\documentclass[pre,twocolumn,preprintnumbers,amsmath,amssymb,nofootinbib,floatfix,superscriptaddress]{revtex4}

\usepackage{graphicx,bm}

\makeatletter

\def\graphicscale{\twocolumn@sw{0.3}{0.4}}
\def\graphicthreescale{\twocolumn@sw{0.3}{0.4}}

\begin{document}

\title{Finite-size scaling at first-order quantum transitions \\
  when boundary conditions favor one of the two phases}

\author{Andrea Pelissetto}
\affiliation{Dipartimento di Fisica dell'Universit\`a di Roma
	``La Sapienza" and INFN, Sezione di Roma I, I-00185 Roma, Italy}

\author{Davide Rossini}
\affiliation{Dipartimento di Fisica dell'Universit\`a di Pisa
        and INFN, Largo Pontecorvo 3, I-56127 Pisa, Italy}

\author{Ettore Vicari} 
\altaffiliation{Authors are listed in alphabetic order.}
\affiliation{Dipartimento di Fisica dell'Universit\`a di Pisa
        and INFN, Largo Pontecorvo 3, I-56127 Pisa, Italy}

\date{\today}

\begin{abstract}
We investigate scaling phenomena at first-order quantum transitions,
when the boundary conditions favor one of the two phases.  We show
that the corresponding finite-size scaling behavior, arising from the
interplay between the driving parameter and the finite size of the
system, is more complex than that emerging when boundary conditions do
not favor any phase.  We discuss this issue in the framework of the
paradigmatic one-dimensional quantum Ising model, along its
first-order quantum transition line driven by an external longitudinal
field.
\end{abstract}

\maketitle


\section{Introduction}
\label{intro}

Zero-temperature quantum phase transitions are phenomena of great
interest~\cite{SGCS-97,Sachdev-book,Vojta-03}. They arise in many-body
systems with competing ground states controlled by nonthermal
parameters.  They are continuous when the ground state of the system
changes continuously at the transition point and correlation functions
develop a divergent length scale. They are instead of first order when
ground-state properties are discontinuous across the transition point.
In general, singularities develop only in the infinite-volume limit.
If the size $L$ of the system is finite, all properties are analytic
as a function of the external parameter driving the
transition. However, around the transition point, low-energy
thermodynamic quantities and large-scale structural properties show a
finite-size scaling (FSS) behavior depending only on the general
features of the transition.  An understanding of these finite-size
properties is important for a correct interpretation of experimental
or numerical data, when phase transitions are investigated in
relatively small systems --- see, e.g.,
Refs.~\cite{Barber-83,Privman-90,GKMD-08,CPV-14,Binder-87,CNPV-14}.
These issues cover a fundamental role also at first-order quantum
transitions (FOQTs), which are very interesting, as they occur in a
large number of quantum many-body systems, such as quantum Hall
samples~\cite{PPBWJ-99}, itinerant ferromagnets~\cite{VBKN-99}, heavy
fermion metals~\cite{UPH-04,Pfleiderer-05,KRLF-09}, etc.

Crossings of the lowest-energy states give rise to FOQTs.  In the
absence of conservation laws, they only occur in the infinite-volume
limit~\cite{clFSS}. In a finite system, the presence of a nonvanishing
matrix element among these states lifts the degeneracy, giving rise to
the phenomenon of avoided level crossing.  The emerging FSS behaviors
have been mostly
investigated~\cite{CNPV-14,CNPV-15,CPV-15,PRV-18b,CPV-15b} assuming
boundary conditions that do not favor any of the different phases at
the FOQT.  In this paper we extend the discussion to boundary
conditions that favor one of the two phases.

We study this issue within the simplest paradigmatic quantum many-body
system, exhibiting a nontrivial zero-temperature behavior: the
one-dimensional quantum Ising chain in the presence of a transverse
field, whose zero-temperature phase diagram presents a line of FOQTs
driven by a longitudinal external field.  Earlier
works~\cite{CNPV-14,CPV-15,PRV-18b,CPV-15b} considered boundary
conditions that are invariant under the ${\mathbb Z}_2$ spin-inversion
symmetry and which therefore do not favor any of the two phases---they
will be called neutral boundary conditions henceforth; for instance,
periodic, antiperiodic, or open boundary conditions (PBC, ABC, and
OBC, respectively).  As we shall see, the FSS emerging when boundary
conditions favor one the two magnetized phases, substantially differs
from, and appears more complex than, those already found for neutral
boundary conditions.  In particular we consider equal fixed boundary
conditions (EFBC), with both boundary states favoring the same
phase. This issue is worth being examined in depth, because boundary
conditions favoring one of the quantum phases at FOQTs are generally
more realistic than the neutral ones such as PBC and ABC.

The paper is organized as follows.  In Sec.~\ref{ismodel} we introduce
the one-dimensional quantum Ising model and some of the observables
which are interesting to be considered along the FOQT line driven by
the longitudinal field. Moreover, we summarize the relevant features
of the FSS behavior in the presence of neutral boundary conditions,
such as PBC, OBC, and ABC.  Sec.~\ref{fbcondfss} reports the main
results of this research, i.e., the numerical study of the finite-size
quantum Ising chain with EFBC favoring one of the two magnetized
phases. Finally, in Sec.~\ref{conclu} we draw our conclusions and
perspectives.

\section{The quantum Ising chain}
\label{ismodel}

\subsection{Model and its observables}
\label{phdia}

The quantum Ising chain in a transverse field is perhaps the simplest
quantum many-body system exhibiting a nontrivial zero-temperature
phase diagram. The corresponding Hamiltonian, in the presence of an
additional longitudinal field, reads
\begin{eqnarray}
H_{\rm Is} = - J \, \sum_{\langle x,y\rangle} \sigma^{(3)}_x \sigma^{(3)}_{y} 
- g\, \sum_x \sigma^{(1)}_x  
- h \,\sum_x \sigma^{(3)}_x \, ,
\label{hedef}
\end{eqnarray}
where ${\bm \sigma}\equiv (\sigma^{(1)},\sigma^{(2)},\sigma^{(3)})$
are the Pauli matrices, the first sum is over all bonds of the chain
connecting nearest-neighbor sites $\langle x,y\rangle$, while the
other sums are over the $L$ sites of the chain.  We assume
$\hslash=1$, $J=1$ and $g>0$.

At $g=1$ and $h=0$, the model undergoes a continuous quantum
transition (CQT) belonging to the two-dimensional Ising universality
class, separating a disordered phase ($g>1$) from an ordered ($g<1$)
one~\cite{Sachdev-book}. For any $g<1$, the field $h$ drives FOQTs
along the $h=0$ line. We are interested in the FSS behavior of the
system along the FOQT line, i.e., in the interplay between the
longitudinal field $h$ and the size $L$, for $g<1$.

In a FOQT, low-energy properties depend on the chosen boundary
conditions, even in the limit $L\to\infty$. If one considers neutral
boundary conditions, the behavior close to the transition can be
completely characterized by considering two magnetized states $| +
\rangle$ and $| - \rangle$ such that~\cite{Pfeuty-70}
\begin{equation}
\langle \pm | \sigma_x^{(3)} | \pm \rangle = \pm \,m_0,\qquad
m_0 = (1 - g^2)^{1/8},
\label{pm0}
\end{equation}
in the infinite-volume limit. Moreover, the longitudinal average
magnetization
\begin{equation}
m(L,h) = \frac{1}{L} \sum_{x} \langle \sigma^{(3)}_{x}\rangle ,
\label{mag}
\end{equation}
is discontinuous, i.e., $\lim_{h\to 0^\pm} \lim_{L\to\infty} m(L,h) = \pm
m_0$.  We should however note that this simple two-level description
does not hold for some other choices of boundary conditions, as we
discuss below.

Here we are going to investigate the
finite-size behavior of the energy difference $\Delta(L,h)$ of the
lowest-energy states,
\begin{equation}
\Delta(L,h) \equiv E_1(L,h) - E_0(L,h),
\label{deltadef}
\end{equation}
the average magnetization $m(L,h)$ defined in Eq.~(\ref{mag}), and the
local magnetization $m_c(L,h)$ at the center of the chain,
\begin{equation}
m_c(L,h) = \langle \sigma^{(3)}_{x_c}\rangle,
\label{mcm}
\end{equation}
where $x_c$ is the central site of the chain (or one of the two
central sites, when $L$ is even).  Let us also introduce the
renormalized average and central magnetizations
\begin{equation}
M = {m\over m_0},\qquad
M_c = {m_c\over m_0},
\label{renmag}
\end{equation} 
which take the values $\pm 1$, in the limit $L \to \infty$,
for $h\to 0^\pm$ and any $g<1$.

The FSS behaviors originating from the cases of neutral boundary
conditions have been already scrutinized in earlier
works~\cite{CNPV-14,CPV-15}.  In order to appreciate the new emerging
features of FSS for boundary conditions favoring one of the two
magnetized phases, it is instructive to first briefly summarize the
known features of FSS for neutral boundary conditions.  This is the
purpose of the remainder of this section.  The next section reports
the results of our analysis, dealing with EFBC favoring one of the two
magnetized phases.

\subsection{Finite-size scaling with periodic and open boundary conditions}
\label{equiopbc}

In a finite system of size $L$ with PBC or OBC, due to tunneling
effects, the lowest eigenstates are superpositions of the states
$|+\rangle$ and $|-\rangle$, defined as $\langle \pm | \sigma_x^{(3)}
| \pm \rangle = \pm \,m_0$.  For $h=0$, their energy difference
\begin{equation}
\Delta_0(L)\equiv \Delta(L,h=0)
\label{delta0l}
\end{equation}
vanishes exponentially as $L$ increases~\cite{Pfeuty-70,CJ-87}:
\begin{eqnarray}
  \Delta_0(L) =  & 2 \, (1-g^2) g^L \, [1+ O(g^{2L})] \;\;
  & \mbox{for OBC} \; , \label{deltaobc} \\
  \Delta_0(L) \approx & 2 \,(\pi L)^{-1/2} (1-g^2) \, g^L \quad \;\;\;
  & \mbox{for PBC} \; . \label{deltapbc}
\end{eqnarray}
On the other hand, the difference $\Delta_{0,i}\equiv E_i-E_0$ for the
higher excited states ($i > 1$) remains finite for $L\to \infty$, in
particular $\Delta_{0,2} = 2(1-g) + O(L^{-2})$ for OBC and
$\Delta_{0,2}=4(1-g)+O(L^{-2})$ for PBC.

The interplay between the size $L$ and the field $h$ gives rise to an
asymptotic FSS of the low-energy properties~\cite{CNPV-14}, in
particular those related to the ground state. The relevant scaling
variable is the ratio between the energy associated with the
longitudinal field $h$, i.e., $2m_0 hL$, and the gap $\Delta_0(L)$ at
$h=0$,
\begin{equation} 
\kappa = {2 m_0 h L\over \Delta_0(L)}.
\label{kappadef}
\end{equation}
The FSS limit corresponds to $L\to \infty$ and $h\to 0$, keeping
$\kappa$ fixed. In this limit, the gap $\Delta$ and the magnetization
behave as
\begin{eqnarray}
&&\Delta(L,h) \approx \Delta_0(L) \, {\cal D}(\kappa),\label{fde}\\
&& M(L,h) \approx M_c(L,h)\approx {\cal M}(\kappa).
\label{efssm}
\end{eqnarray}  
Note---this remark will be important in the case of EFBC---that the
energy $2 m_0 h L$ can be interpreted as the difference between the
magnetic energy at the given value of $h$ and that at the value where
the gap displays its minimum, i.e., for $h=0$.

In the PBC and OBC cases, the scaling functions can be exactly
computed.  Since, close to the FOQT, the low-energy spectrum is
characterized by the crossing of the two lowest levels, while the
energy differences $\Delta_{0,i}$ with the other ones remain finite,
FSS functions can be obtained by performing a two-level
truncation~\cite{CNPV-14,PRV-18}, keeping only the two lowest energy
levels $|\pm \rangle$.  Then, a straightforward calculation leads to
the scaling functions
\begin{eqnarray}
{\cal D}(x)={\cal D}_{2l}(x) = \sqrt{1 + x^2}, \label{fdm}\\
{\cal M}(x)={\cal M}_{2l}(x) = {x \over \sqrt{1+ x^2}}. \label{fmm}
\end{eqnarray}  
It is important to note that, in the derivation of Eq.~(\ref{fmm}), we
have assumed that the magnetization $M$ of the two states $|\pm
\rangle$ is $\pm 1$, respectively.

It is also worth mentioning that an analogous FSS behavior emerges
when, instead of the homogeneous field $h$, we consider an external
longitudinal field $h_l$ applied at one site
only~\cite{CNPV-14,PRV-18}. The only difference amounts to replacing
the product $h L$ with $h_l$ in the definition (\ref{kappadef}) of the
scaling variable $\kappa$, while the two-level truncation holds as
well.  In a sense, the system behaves rigidly at the FOQT when PBC or
OBC are considered, i.e.~its response to global or local longitudinal
perturbations is analogous.

\subsection{Finite-size scaling with antiperiodic  boundary conditions}
\label{equiabc}

The size dependence of the gap $\Delta_0(L)$ at FOQTs may
significantly depend on the boundary conditions, exhibiting a
power-law behavior in some
cases~\cite{CJ-87,LMMS-12,CNPV-14,CPV-15}. For example, for
ABC~\cite{CPV-15} we have
\begin{equation}
\Delta_0(L) \equiv \Delta(L,h=0)=
{g\over 1-g} \, {\pi^2\over L^2} + O(L^{-4}).
\label{deltaabc}
\end{equation}
This is related to the fact that the low-energy states for ABC are
one-kink states (characterized by a nearest-neighbor pair of
antiparallel spins), which behave as one-particle states with
$O(L^{-1})$ momenta.  Therefore, using Eq.~(\ref{kappadef}), we expect
FSS to hold if we define the scaling variable $\kappa$ as
\begin{equation}
\kappa\sim h L^3 .
\label{kappaabc}
\end{equation}
Such a behavior has been indeed observed in
Ref.~\cite{CNPV-14}. However, in this case, scaling functions cannot
be obtained by performing a two-level truncation, because the
low-energy spectrum at the transition point presents a tower of
excited stated with $\Delta_{0,i}=O(L^{-2})$, at variance with the OBC
and PBC case, where only two levels matter, close to the transition
point.

Note that a similar behavior also emerges for fixed and opposite
boundary conditions~\cite{CPV-15}, for which the lowest states can be
associated with kink states, as well.

\section{Boundary conditions favoring one of the two phases}
\label{fbcondfss}

Let us now focus on a quantum Ising chain of size $L$ with EFBC
favoring one of the two magnetized phases. We consider $L+2$ spins
defined at the lattice sites $x=0,\ldots,L+1$ and the Hamiltonian
\begin{eqnarray}
H_{\rm Is} = - \sum_{x=0}^{L}  \,\sigma^{(3)}_x \sigma^{(3)}_{x+1} 
- g\, \sum_{x=1}^L \sigma^{(1)}_x  
- h \,\sum_{x=1}^L \sigma^{(3)}_x \,.
\label{hedef2}
\end{eqnarray}
EFBC are fixed by restricting the Hilbert space to states $|s\rangle$
such that $\sigma^{(3)}_0 |s\rangle = - |s\rangle$ and
$\sigma^{(3)}_{L+1} |s\rangle = - |s\rangle$.

As we shall see below, the interplay between the size $L$ and the bulk
longitudinal field $h$ gives rise to a more complex finite-size
behavior, with respect to that of neutral boundary conditions.  In the
following, this issue is investigated by analyzing numerical results
for two values of $g$, i.e., $g=0.5$ and $g=0.8$, obtained by exact
diagonalization, up to $L \approx 22$, and density-matrix
renormalization group (DMRG) methods~\cite{DMRG} for larger sizes, up
to $L \approx 300$~\cite{footnote}.

EFBC can be naturally enforced with exact methods. On the other hand,
DMRG naturally works with OBC.  Therefore, in this case we effectively
simulated a nonhomogeneous chain of $L+2$ sites with OBC, and then
added two large local magnetic fields on the first and last site,
whose net effect is that of removing from the low-energy spectrum the
unwanted states, i.e., those corresponding to $|\!\!\uparrow\rangle$
occupancies in the two boundary sites~\cite{footnote}.

\begin{figure}[!t]
  \includegraphics[width=0.95\columnwidth]{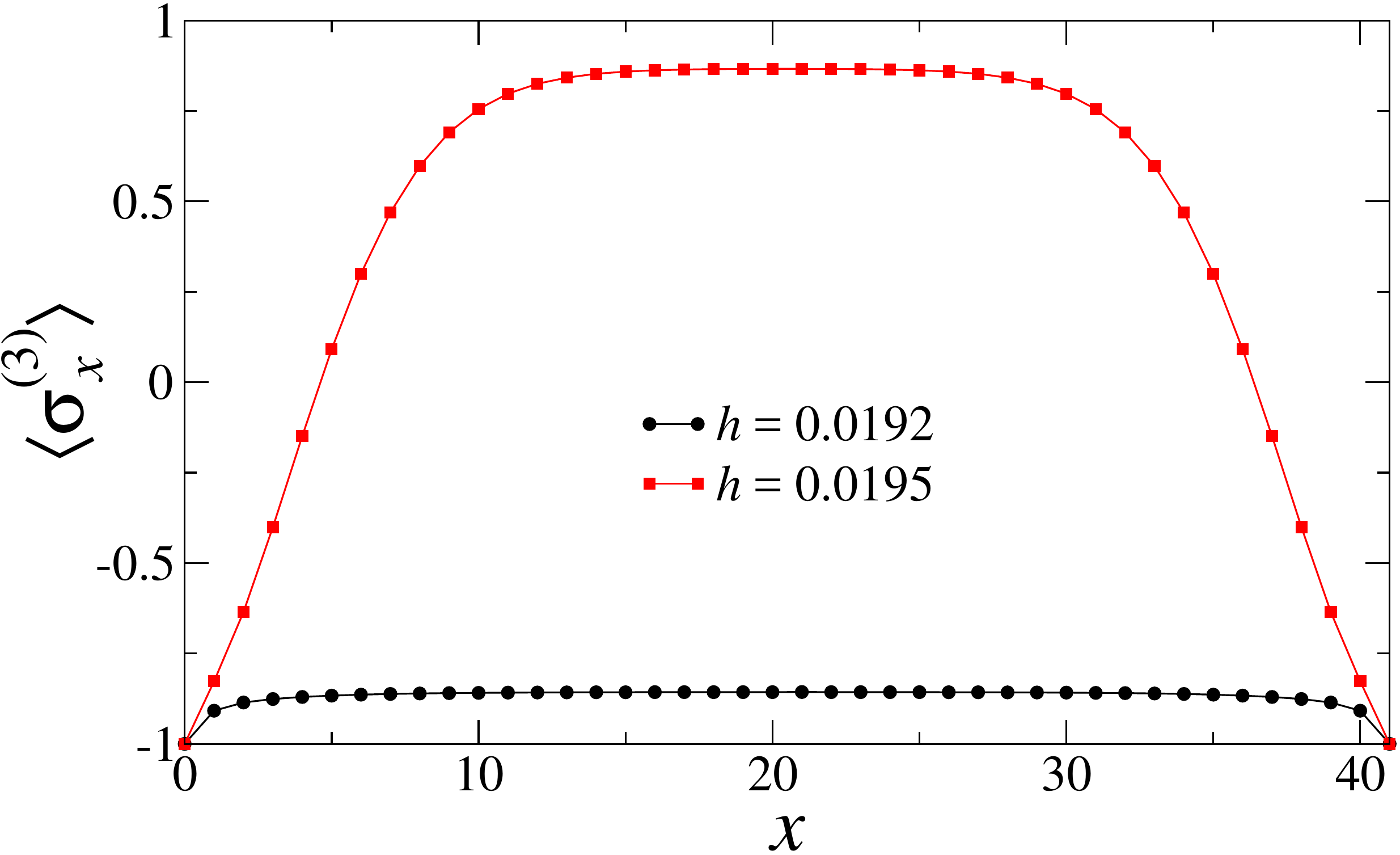}
  \caption{Magnetization profile along the magnetic-field direction in
    the presence of EFBC, for $L=40$, $g=0.8$, and two values of the
    longitudinal field $h$.  Note that the local magnetization for
    $x=0$ and $x=L+1=41$ is exactly $-1$, because of the boundary
    conditions.}
    \label{fig:Mprofile}
\end{figure}

Before entering the details of our discussion, let us provide a
qualitative picture of the system's response to the longitudinal field
$h$, by analyzing the magnetization profile.
Figure~\ref{fig:Mprofile} highlights the net macroscopic effect of two
values of $h$, on a system with $L=40$ sites, $g=0.8$, and EFBC. It
emerges that, if the longitudinal field is not sufficiently strong,
the magnetization profile displays a nearly flat behavior: because of
the boundary conditions, the value of $\langle \sigma^{(3)}_x \rangle$
always stays close to $-1$, corresponding to the
$|\!\!\downarrow\rangle$ state on each site $x$ of the chain.
Conversely, a sufficiently large value of $h$ is able to substantially
modify the profile, inducing a local magnetization in the bulk of the
chain which is close to $+1$ (corresponding to a $|\!\uparrow\rangle$
state), thus opposed to that favored by the EFBC.  Thus, there should
exist some threshold value (or region) of $h$, separating the two
distinct behaviors.

\begin{figure}[!t]
  \includegraphics[width=0.95\columnwidth]{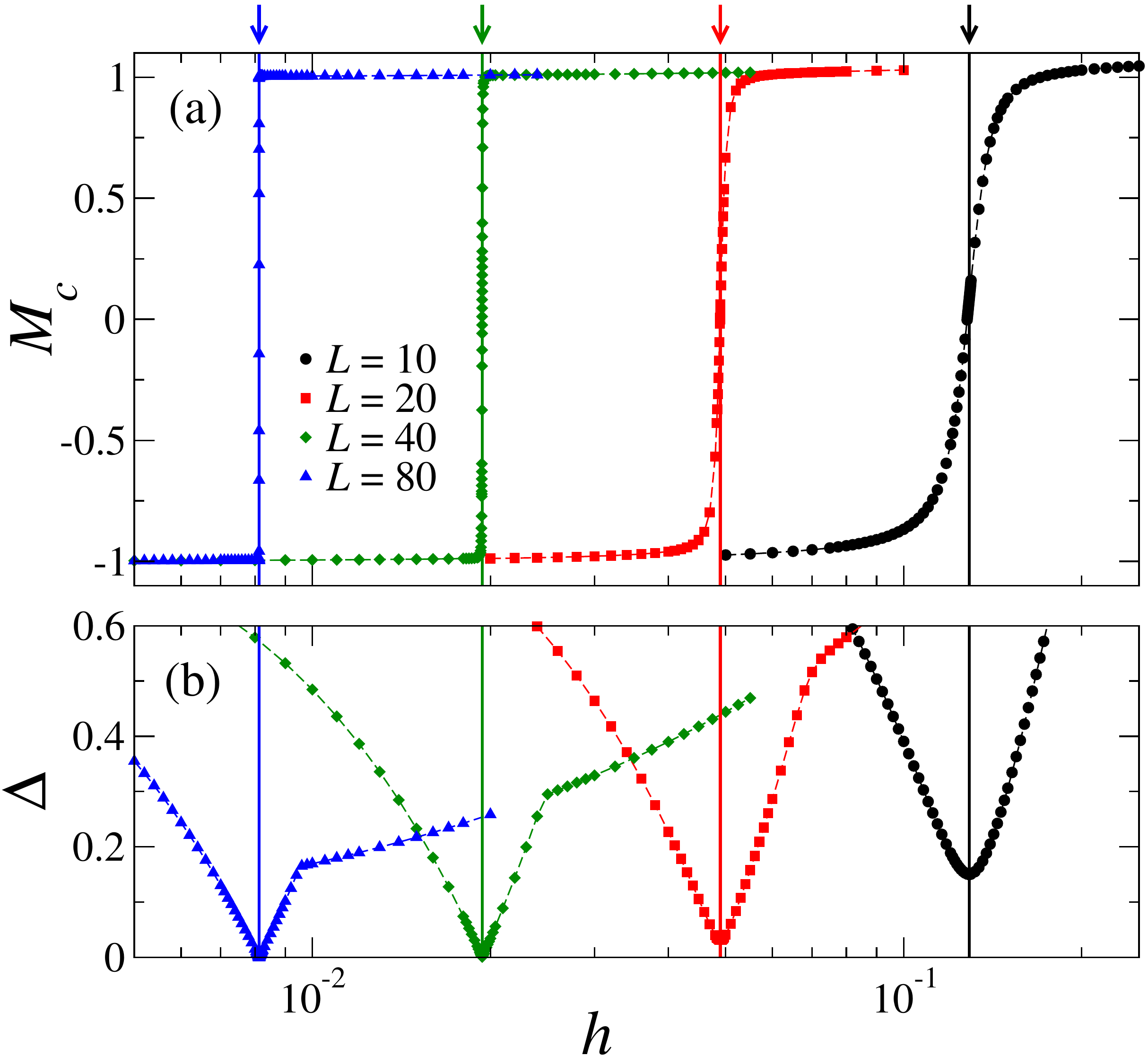}
  \caption{Central magnetization $M_c$ [panel (a)] and energy gap
    $\Delta$ [panel (b)] in the Ising model with EFBC, as a function
    of the longitudinal field $h$, for fixed transverse field $g=0.8$.
    The various data sets correspond to different system sizes, as
    indicated in the legend.  The continuous vertical lines and arrows
    denote the magnetic fields $h_{tr}(L)$, which correspond to the
    minimum of the energy gap, as displayed in panel (b).}
  \label{fig:Mc}
\end{figure}

A glimpse at the numerical data of the central magnetization $M_c$
versus $h$ presented in Fig.~\ref{fig:Mc}(a) immediately spotlights
that the above outlined transition from $M_c\approx -1$ to $M_c\approx
+1$ is indeed very rapid, when increasing $h$ across an $L$-dependent
value $h_{tr}(L)>0$ (indicated with arrows and continuous vertical
lines in the figure), which approaches $h=0$ when increasing $L$.
While for small sizes we observe a smooth crossover between the two
phases, this crossover becomes sharper when increasing $L$, until we
are not able to distinguish the transition region (see the results for
$L=80$ in the figure).  The transition region also shrinks if we fix
$L$ and consider a transverse field $g$ farther from $g=1$.  This
sharp crossover corresponds to the minimum $\Delta_m(L)$ of the energy
difference of the two lowest states, see Fig.~\ref{fig:Mc}(b), evidencing
the correspondence with an avoided level crossing, where the energies
of such two states get closer and closer with increasing $L$.
Actually, we may define $h_{tr}(L)$ as the value of $h$ where the gap
shows its minimum, $\Delta_m(L) \equiv \Delta [L,h_{tr}(L)]$,
which vanishes in the thermodynamic limit $L \to \infty$.

We stress that the minimum of the energy gap, that we can interpret as
the finite-size pseudotransition point, is located at $h_{tr}(L)>0$,
at variance with what occurs for neutral boundary conditions, where
the minimum is always located at $h=0$ for any value of $L$.

Summarizing, we can identify three distinct regions, corresponding to:
$a)$ small values of $h$, where the longitudinal field ($h < h_{tr}$)
is unable to modify the phase of the system stabilized by the EFBC;
$b)$ intermediate values of $h$ around $h_{tr}(L)$, where the
interplay between the boundary conditions and the field induces a
sharp transition; $c)$ large values of $h$, where the bulk phase is
determined by the field ($h > h_{tr}$).  Below we separately discuss
the emerging physics in these three regions.

\subsection{Small-$h$ region}
\label{smallh}

\begin{figure}[!t]
  \includegraphics[width=0.95\columnwidth]{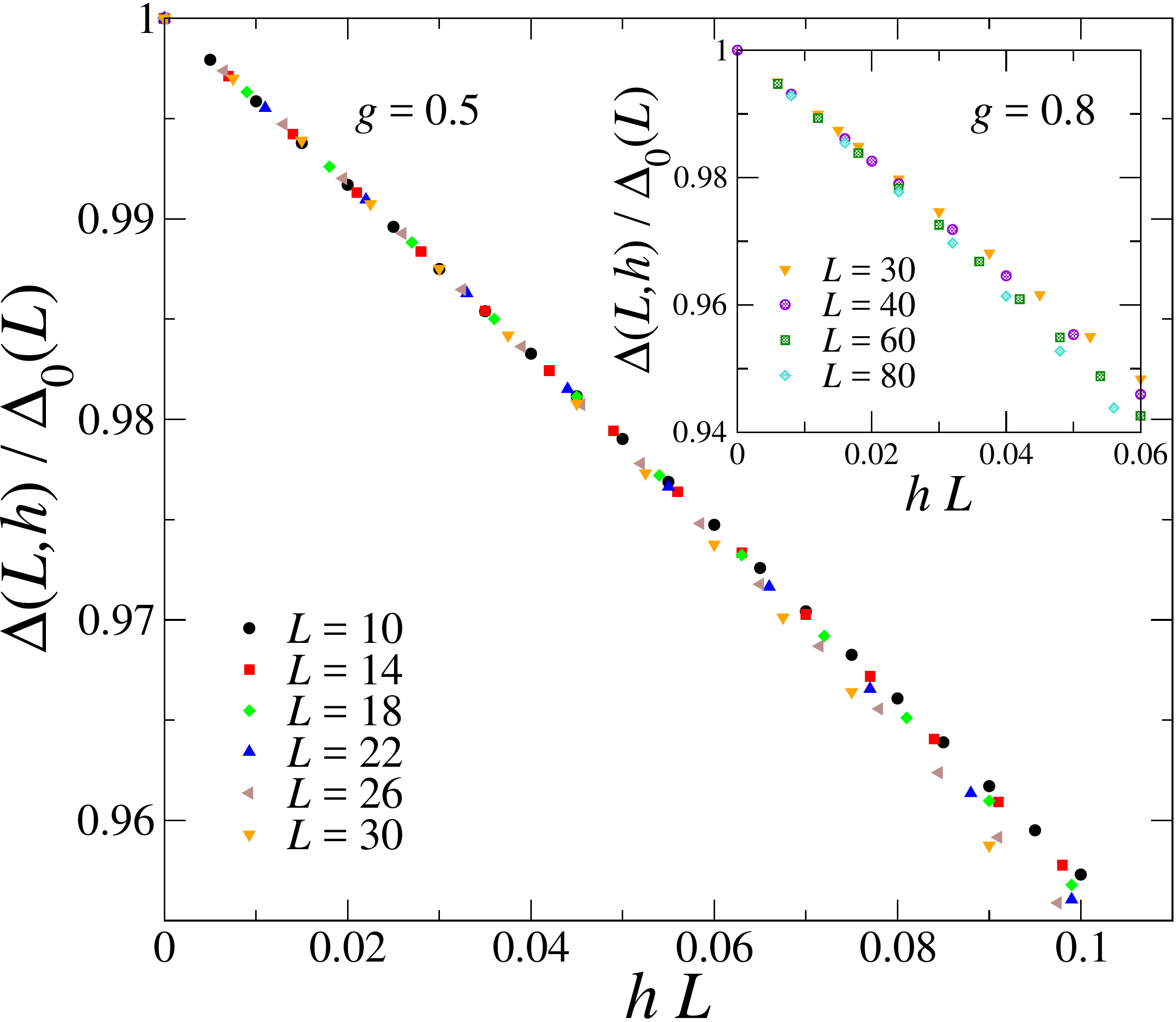}
  \caption{Energy gap ratio $\Delta(L,h)/\Delta_0(L)$ as a function
    of the rescaled variable $hL$, for $g=0.5$ (main frame) and
    for $g=0.8$ (inset).  The symbols correspond to different values of
    $L$. }
  \label{fig:Gaps_h0}
\end{figure}

For $h=0$, the phase with negative magnetization is favored by the
boundary conditions. Correspondingly, we have
\begin{equation}
\lim_{L\to\infty} M(L,0) = \lim_{L\to\infty} M_c(L,0) = - 1,
\label{mh0}
\end{equation}
and the gap is finite~\cite{CPV-15b}:
\begin{equation}
\Delta_0(L) = 4 (1-g) + O(L^{-2}).
\label{deltal0f}
\end{equation}
The finite-size transition to the phase with positive magnetization
occurs for $h \approx h_{tr}(L)>0$. Therefore, at fixed $L$, we expect
observables to be smooth for $h\approx 0$ and to have a regular
expansion around $h=0$, which is predictive up to $h_{tr}(L)>0$. The
$L$-dependence of the observables in this regime depends on their
nature.  Local observables that are defined far from the boundaries
(they are localized in a region whose distance from the boundaries is
much larger than the correlation length $\xi$ of the system) are
expected to have a negligible dependence on $L$, thus they smoothly
depend only on $h$, for $h < h_{tr}(L)$.  In particular, this is the
case for the central magnetization $M_c$: when $L\gg \xi$ and
$h<h_{tr}(L)$, $M_c\approx f_m(h)$, with a little dependence on $L$.
This agrees with our numerical findings (data not shown here).  On the
other hand, the energy gap, which is a global quantity, obeys the
scaling relation
\begin{equation}
\Delta(L,h) \approx \Delta_0(L) \, f_\Delta(hL),
\label{deltah0f}
\end{equation}
where $f_\Delta(x)$ is a smooth function of $x$, around $x=0$.  This
is shown in Fig.~\ref{fig:Gaps_h0}, where we clearly observe data
collapse for $g=0.5$ already at small sizes (main frame), while for
$g=0.8$ the collapse occurs at larger sizes (inset).  The slower
approach to scaling when raising $g$ along the FOQT line can be easily
explained by the increasing correlation length $\xi\sim (1-g)^{-1}$,
when approaching the CQT at $g=1$.

\subsection{Transition region}
\label{trreg}

\begin{figure}[!t]
  \includegraphics[width=0.95\columnwidth]{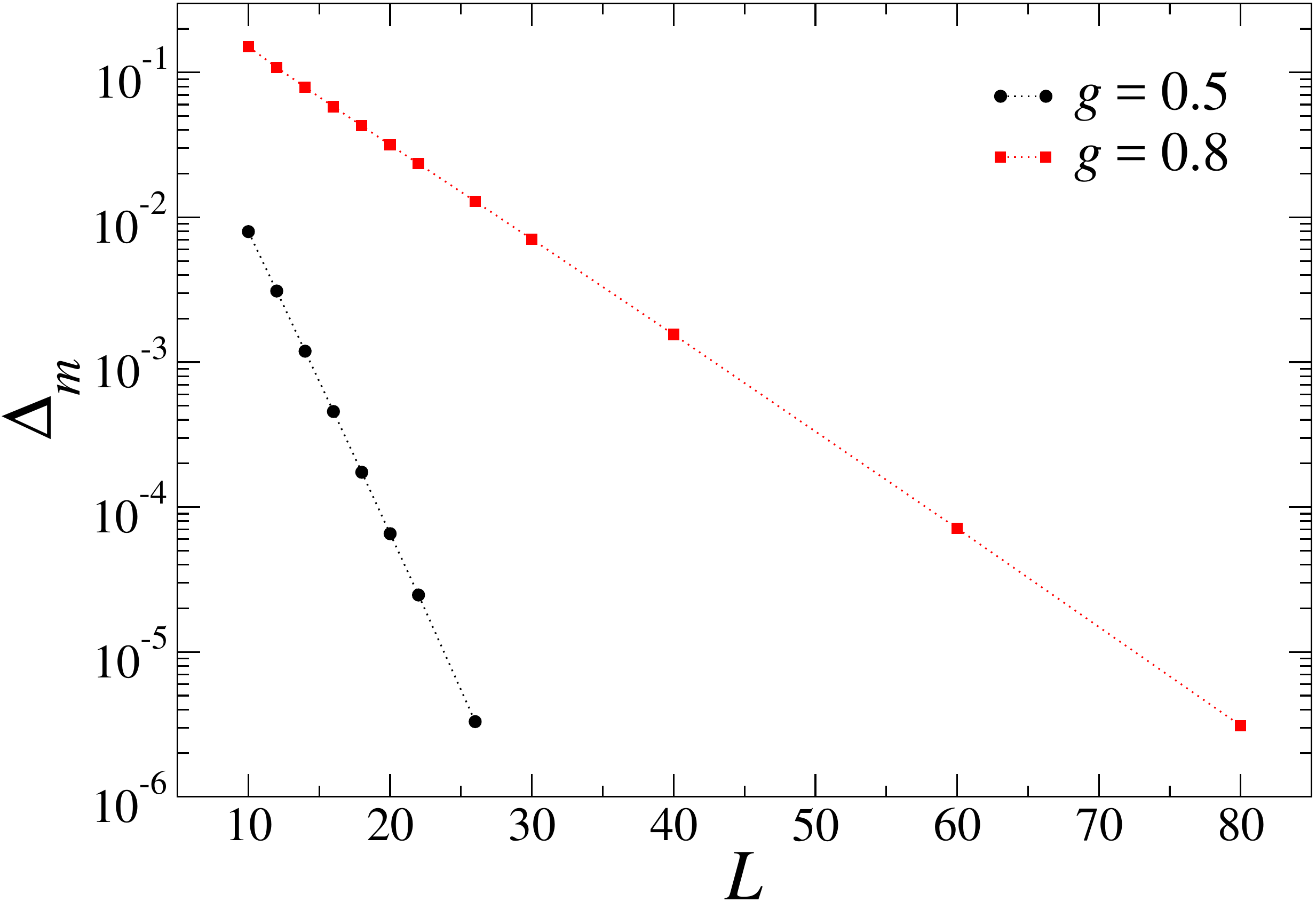}
  \caption{Minimum gap $\Delta_m$ for $h = h_{tr}(L)$ as a function of the
    system size $L$, for two different values of $g$, as explained in
    the legend.  The dotted lines are only meant to guide the eye:
    they show that the data approximately behave as $\Delta_m\sim
    e^{-b L}$.}
  \label{fig:Delta_min}
\end{figure}

For $h\approx h_{tr}(L)$, the gap between the ground state and the
first excited state becomes small.  We were able to reliably determine
its minimum $\Delta_m(L)$ up to values of the order of $10^{-6}$,
within the accuracy of our numerical simulations --- see
Fig.~\ref{fig:Delta_min}.  Our data show that $\Delta_m(L) \sim
e^{-bL}$ for $L$ sufficiently large.  We estimate $b\approx 0.481$ and
$b\approx 0.15$, for $g=0.5$ and $g=0.8$, respectively.

\begin{figure}[!t]
  \includegraphics[width=0.95\columnwidth]{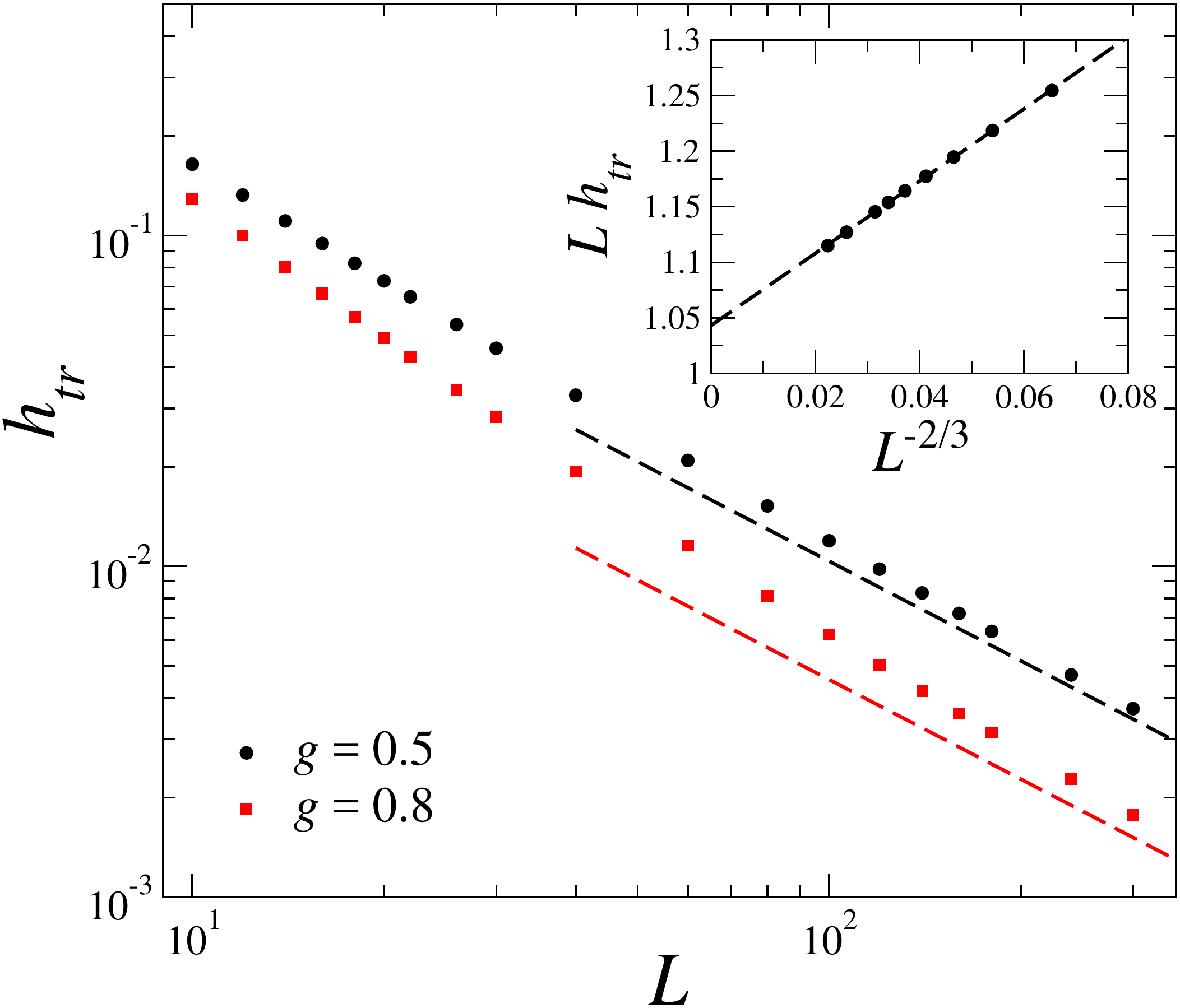}
  \caption{Longitudinal-field value $h_{tr}(L)$ at which the system
    exhibits a minimum in the gap, as a function of the size $L$, for
    two different values of $g$, as explained in the legend.  The
    dashed lines correspond to the asymptotic behavior
    $h_{tr}(L)\approx \eta/L$, cf.~Eq.~(\ref{asyhtr}).  The inset
    displays $L \, h_{tr}(L)$ for $g=0.5$, as a function of
    $L^{-2/3}$: data fall on a straight line, showing the presence of
    corrections of order $L^{-2/3}$.  The uncertainty on the estimates
    of $h_{tr}(L)$ is always smaller than the symbols sizes.}
  \label{fig:Delta_pos}
\end{figure}

The position of $h_{tr}(L)$ can be determined more accurately, since
its estimate does not require to probe regions with very small gaps,
where numerical methods, such as Lanczos or DMRG, may typically
encounter problems.  Using DMRG, we obtained results up to $L=300$, as
shown in Fig.~\ref{fig:Delta_pos}. The resulting estimates of
$h_{tr}(L)$ are very accurate. In particular, the relative accuracy is
of the order of $10^{-6}$ for $g= 0.5$ ($L\gtrsim 100$) and of
$10^{-4}$ for $g=0.8$.  Results are consistent with the expected $1/L$
asymptotic behavior, i.e., with
\begin{equation}
\lim_{L\to\infty} L \, h_{tr}(L) = \eta \, ,
\label{asyhtr}
\end{equation}
where $\eta$ decreases with $g$.  However, the next-to-leading
corrections are not consistent with the expected analytic $1/L^2$
behavior.  Our data indicate $L\,h_{tr}(L) = \eta + O(L^{-\zeta})$,
where $\zeta$ is an exponent that is strictly less than 1 (see the
inset of Fig.~\ref{fig:Delta_pos}).  For both $g=0.5$ and $g=0.8$, the
data for the largest sizes are consistent with the ansatz
\begin{equation}
L\, h_{tr} = \eta + a_1 L^{-\zeta} + a_2 L^{-1}
\label{fithl}
\end{equation}
with $\zeta\approx 2/3$ (the $L^{-1}$ term represents an analytical
corrections which must be generally present).  If we fit our data for
$g=0.5$ and $L\ge 100$ to the ansatz~(\ref{fithl}), we obtain $\eta =
1.0370(5)$, $a_1=4.0(1)$, and $\zeta=0.67(1)$ with $\chi^2/{\rm
  dof}\approx 1$ (dof is the number of degrees of freedom of the fit,
and $\chi^2$ is the sum of the residuals). Errors are estimated by
also taking into account the variation of the results with the minimum
size allowed in the fit.  Analogously, for $g=0.8$, fitting the
available data for $L\gtrsim 60$ gives $\eta = 0.455(5)$,
$a_1=2.6(6)$, and $\zeta=0.64(4)$, with $\chi^2/{\rm dof}\lesssim
0.5$.  We will return to this point later, in Sec.~\ref{lhreg},
providing an explanation for the $O(L^{-\zeta})$ correction with
$\zeta=2/3$ in Eq.~(\ref{fithl}).

The above picture, and in particular the asymptotic behavior of
$h_{tr}(L)$, is also supported by the analysis of the $g\to 0$ limit,
where the energy levels of model (\ref{hedef2}) can be easily
computed, obtaining that $h_{tr}(L) = 2/L$, thus $\eta=2$, in the
limit $g\to 0$. In this limit, it is also trivial to verify that that
gap $\Delta_m$ at $h_{tr}(L)$ decreases exponentially as $e^{-b L}$
with $b \approx -\ln g$.

For $h$ close to $h_{tr}(L)$, we can define a FSS in terms of the
scaling variable
\begin{equation}
y = {2 m_0 L \, [h - h_{tr}(L)]\over \Delta_m(L)} \, .
\label{wdef}
\end{equation}
This variable is the analogue of $\kappa$ defined in
Eq.~(\ref{kappadef}).  The essential difference is related to the fact
that the finite-size pseudotransition occurs at $h = h_{\rm tr}(L)$,
and not at $h = 0$.  Therefore, the relevant magnetic energy scale is
the difference between the magnetic energy at $h$ and that at $h_{\rm
  tr}(L)$, while the relevant gap is the one at $h_{tr}(L)$. The
infinite-volume critical point $h=0$ lies outside the region in which
FSS holds. Note that a crucial point in the definition of the scaling
variable $y$ is that the values of $h_{tr}(L)$ and $\Delta_m(L)$ must
be those associated with the minimum of the gap for the given size
$L$, i.e., they cannot be replaced with their asymptotic behaviors.

\begin{figure}[!t]
  \includegraphics[width=0.95\columnwidth]{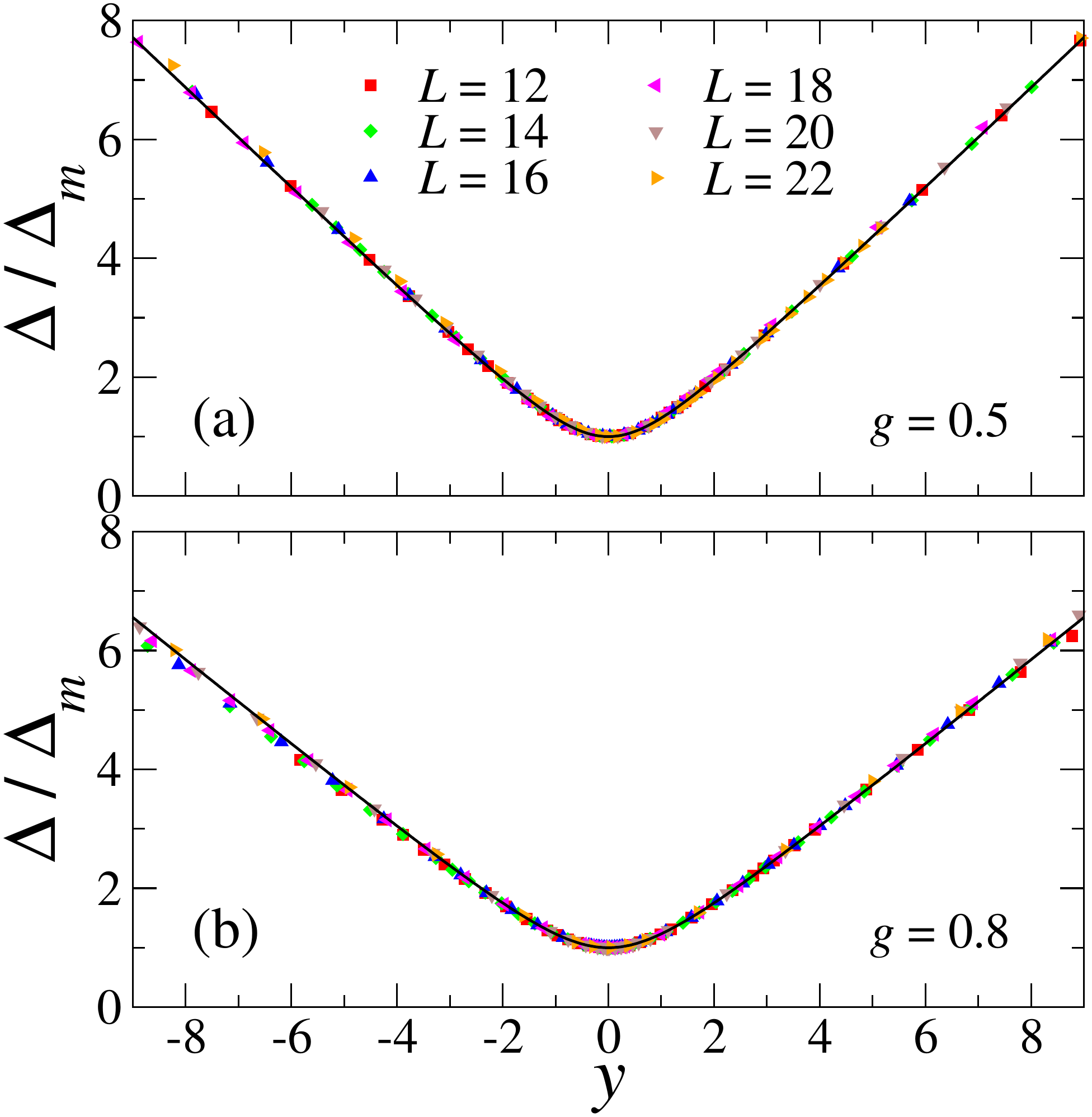}
  \caption{Ratio $\Delta/\Delta_m$, where $\Delta\equiv \Delta(L,h)$
    is the gap and $\Delta_m \equiv \Delta(L,h_{tr})$, as a function
    of the scaling variable $y$ defined in Eq.~\eqref{wdef}, for
    several system sizes (see the legend).  Panel (a) is for $g=0.5$,
    while panel (b) is for $g=0.8$.  We also report (continuous black
    curve) the two-level prediction \eqref{fdmy}.}
  \label{fig:Gap_2lev}
\end{figure}

\begin{figure}[!t]
  \includegraphics[width=0.95\columnwidth]{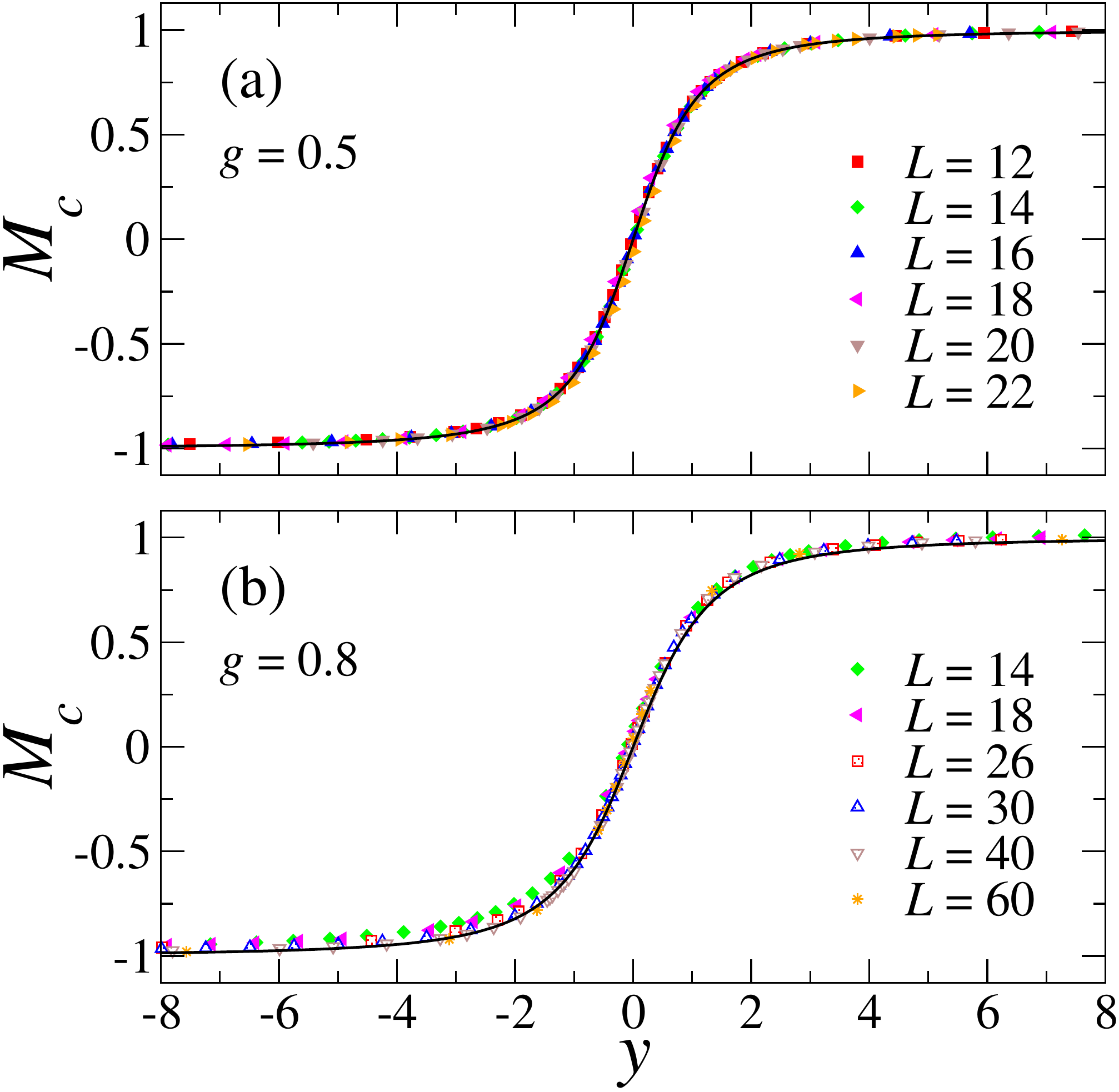}
  \caption{Local magnetization at the center of the chain $M_c$, as a
    function of the scaling variable $y$, for several system sizes
    (see the legend).  Panel (a) is for $g=0.5$, while panel (b) is
    for $g=0.8$.  We also report (continuous black curve) the
    two-level prediction \eqref{twolfbc-mc}, with the same
    constant $c$ as in Fig.~\ref{fig:Gap_2lev}.}
  \label{fig:Mc_2lev}
\end{figure}

\begin{figure}[!t]
  \includegraphics[width=0.95\columnwidth]{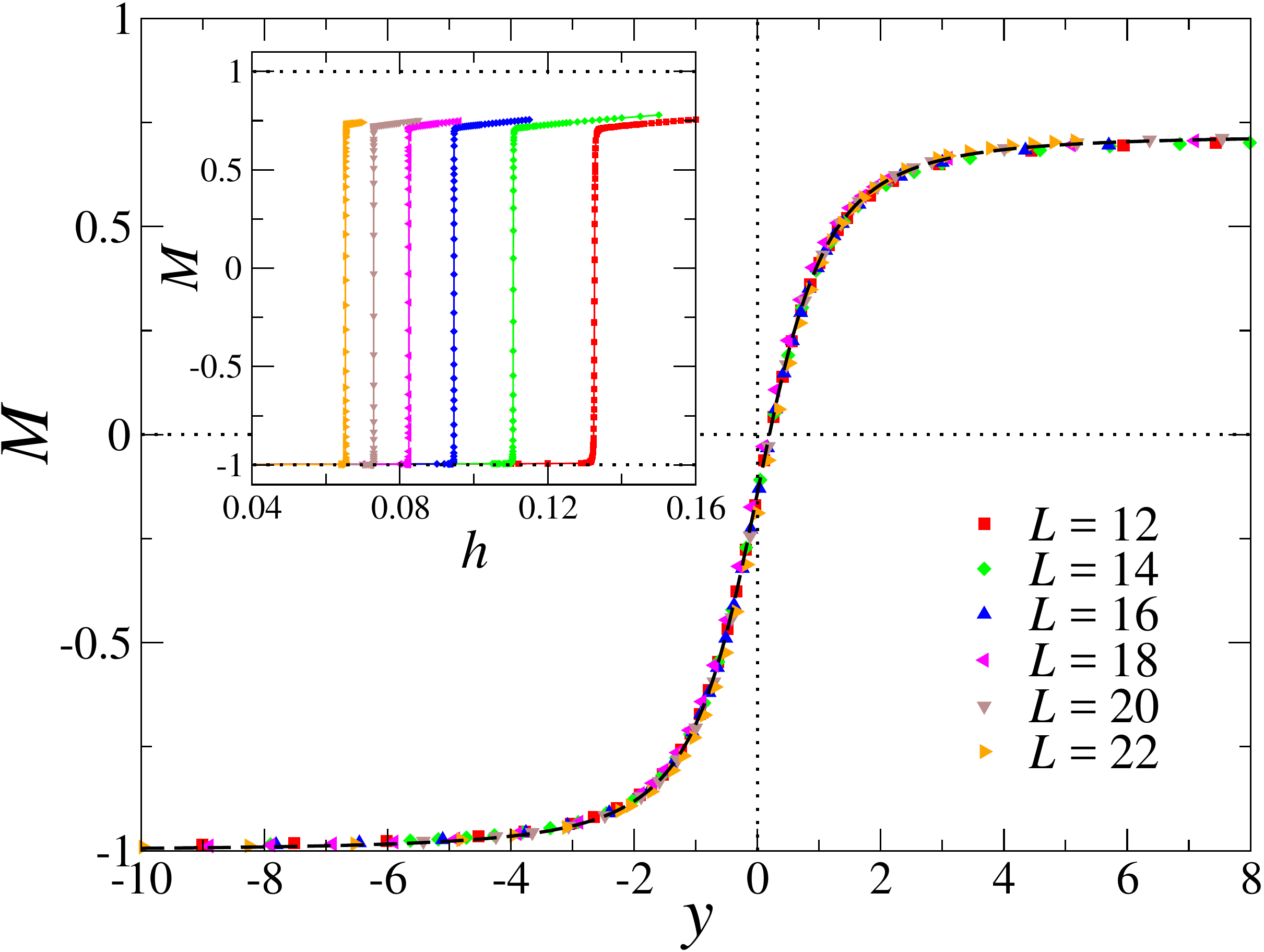}
  \caption{Average magnetization $M$, as a function of the scaling
    variable $y$, for several system sizes (see the legend) and $g =
    0.5$.  We also report (continuous black curve) the two-level prediction
    \eqref{twolfbc-m}, where  the constant $c$ is the same as in
    Fig.~\ref{fig:Gap_2lev}.  In the inset we report $M$ as a function of $h$.}
  \label{fig:Maverage_2lev}
\end{figure}

For $h\approx h_{tr}(L)$, observables are expected to develop a FSS
behavior given by
\begin{eqnarray}
\Delta(L,h) & \approx & \Delta_m(L)\, {\cal D}_f(y) \,,   \label{deltasca} \\
M_c(L,h) & \approx & {\cal M}_{cf}(y) \,,   \label{mcsca} \\
M(L,h) & \approx & {\cal M}_f(y) \,.   \label{msca}
\end{eqnarray}
These predictions are nicely supported by the data, see
Figs.~\ref{fig:Gap_2lev}, ~\ref{fig:Mc_2lev}, and
\ref{fig:Maverage_2lev}.  We observe that the convergence to the
asymptotic infinite-volume limit appears to be slightly slower for
$g=0.8$, as is reasonable for values of the transverse field which are
closer to $g=1$ [this is especially evident in panel (b) of
  Fig.~\ref{fig:Mc_2lev}, for the local magnetization at the center of
  the chain].

Note that, close to $h_{tr}(L)$, there are only two relevant levels
(those whose energy difference becomes exponentially small) and
therefore we can again apply a two-level truncation of the state space
to compute the FSS functions. For the energy gap we recover
Eq.~\eqref{fdm}, apart from a trivial but unique renormalization of
the argument, $y = c \,x $, i.e.
\begin{equation}
 {\cal D}_f(y) = {\cal D}_{2l}(y/c),
\label{fdmy}
 \end{equation}
where ${\cal D}_{2l}$ is the function obtained by the two-level
truncation, cf.~Eq.~(\ref{fdm}).  This is once more supported by the
data shown in Fig.~\ref{fig:Gap_2lev}, which nicely fit the two-level
scaling behavior (with $c\approx 1.18$ for $g=0.5$, and $c \approx
1.4$ for $g=0.8$).  This confirms that, even for EFBC, the FOQT is
characterized by the crossing of two quantum energy levels.

For the magnetization we should be more careful, as Eq.~\eqref{fmm}
has been derived under the assumption that the magnetization of the
two states is $\pm 1$, respectively. For the central magnetization,
this assumption is satisfied, as can be seen from
Fig.~\ref{fig:Mc_2lev} [indeed ${\cal M}_{cf}(y)$ converges to $\pm
  1$, as $y\to \pm \infty$], and hence we expect
\begin{equation}
 {\cal M}_{cf}(y) = {\cal M}_{2l}(y/c) ,
 \label{twolfbc-mc} 
 \end{equation}
with ${\cal M}_{2l}$ given in Eq.~(\ref{fmm}), and the same constant
$c$ obtained in the analysis of the gap. The results shown in
Fig.~\ref{fig:Mc_2lev} are fully consistent.  On the other hand, the
assumption is not true for the average magnetization $M$. Indeed, for
$y\to \infty$, it converges to a value that is less than 1 --- see the
inset of Fig.~\ref{fig:Maverage_2lev}. We can identify this value with
the magnetization $M_{2s}$ of the ground state that is obtained by
approaching $h_{tr}(L)$ from above, and is the relevant one in the
limit $y\to\infty$.  Numerically, we find $M_{2s} \approx 0.72$ and
$M_{2s} \approx 0.52$ for $g=0.5$ and $g=0.8$, respectively.  Using
the two-level truncation, we predict for the scaling function
\begin{eqnarray}
{\cal M}_f(y) & = & {\cal M}_{2l,a}(y/c), \nonumber  \\
{\cal M}_{2l,a}(x) & = & 
   {M_{2s} - 1\over 2} + {M_{2s} + 1\over 2} {x\over \sqrt{1 + x^2} },
\label{twolfbc-m}
\end{eqnarray}
which interpolates between ${\cal M}_{2l,a}(x\to -\infty)=-1$
and ${\cal M}_{2l,a}(x\to \infty)=M_{2s}$.
Numerical results are in perfect agreement --- see Fig.~\ref{fig:Maverage_2lev} 
for the data at $g=0.5$.

It is important to note that the energy difference between the ground
state and the higher excited states is expected to be of order $h$,
hence of order $1/L$ at the transition point.  The presence of this
tower of states does not contradict the validity of the two-level
approximation, since the relevant ratios $\Delta(L)/\Delta_{0,n}(L)$
vanish exponentially for all $n\ge 2$.

An interesting question concerns the nature of the two states which
give rise to the above level-crossing scenario in the large $L$
limit. One of them is the ground state for $h=0$, i.e., the negatively
magnetized state with $M = -1$ in the large-$L$ limit; the other one
is a state with a large positively magnetized region around the
center, and two negatively magnetized regions at the boundaries,
separated by a kink and an antikink close to the left and right
boundary, respectively --- see, e.g., Fig.~\ref{fig:Mprofile}. Close
to the transition, the size of such regions at the boundaries must be
of order $L$, to guarantee that the average magnetization is strictly
less than 1.

\subsection{Large-$h$ region}
\label{lhreg}

\begin{figure}[!t]
  \includegraphics[width=0.95\columnwidth]{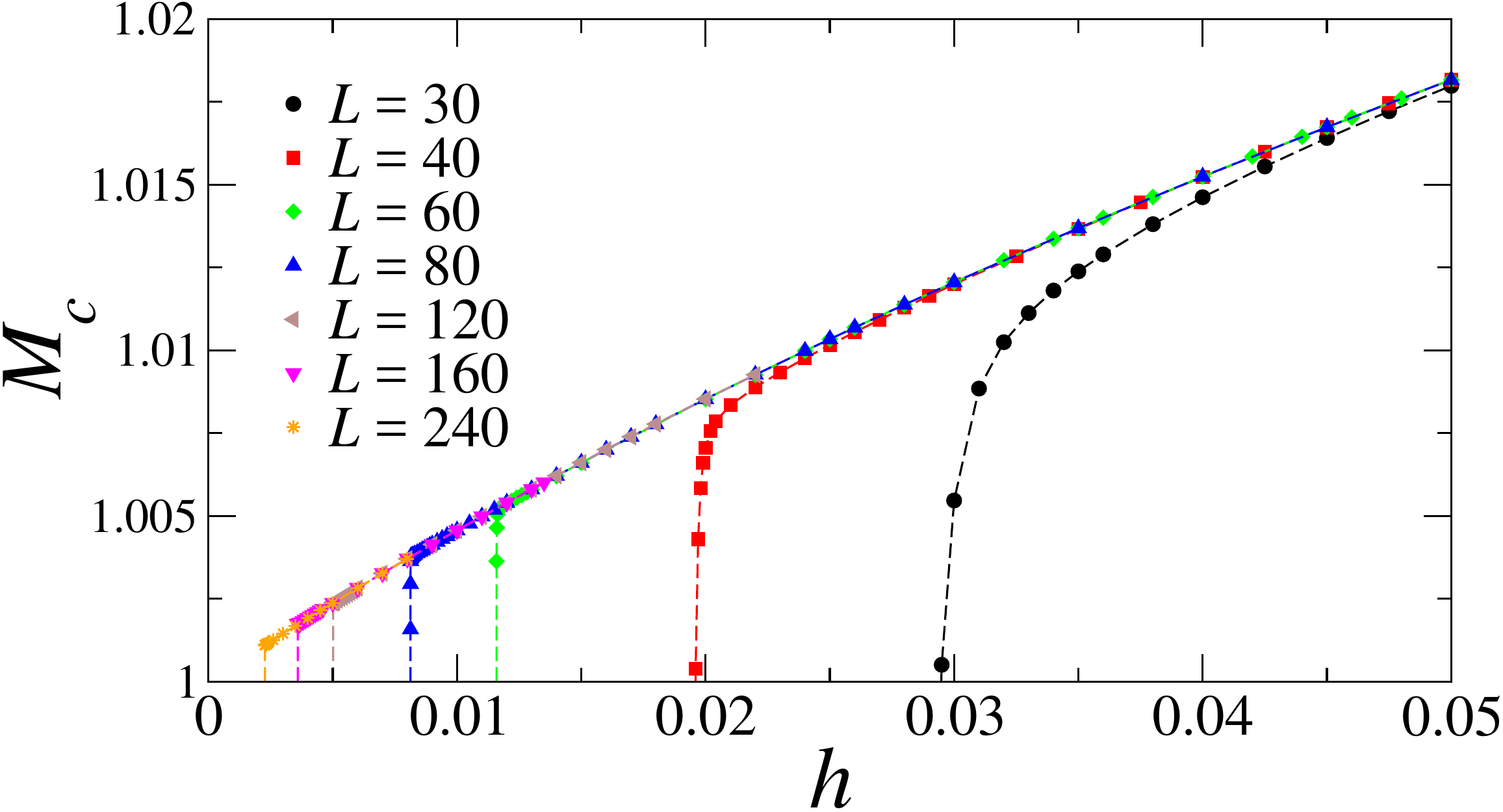}
  \caption{The local central magnetization $M_c$ in the large-$h$
    region, for fixed $g=0.8$ and different system sizes, as indicated
    in the legend.}
  \label{fig:largehMc}
\end{figure}

We now discuss the main features of the finite-size behavior for
$h>h_{tr}$. As stated above, in this regime, low-energy states are
characterized by a positively magnetized region around the center of
the chain and by two negatively magnetized regions at the
boundaries. The nature of the central region can be easily understood
by considering the central magnetization $M_c$, displayed in
Fig.~\ref{fig:largehMc}. As $L$ increases, results rapidly approach a
function of $h$ only. This trivial dependence on $h$ is expected,
since $M_c$ is a local quantity which is not sensitive to the
boundaries.  On the other hand, as we shall see below, the behavior in
proximity of the boundaries is more involved.

\begin{figure}[!t]
  \includegraphics[width=0.95\columnwidth]{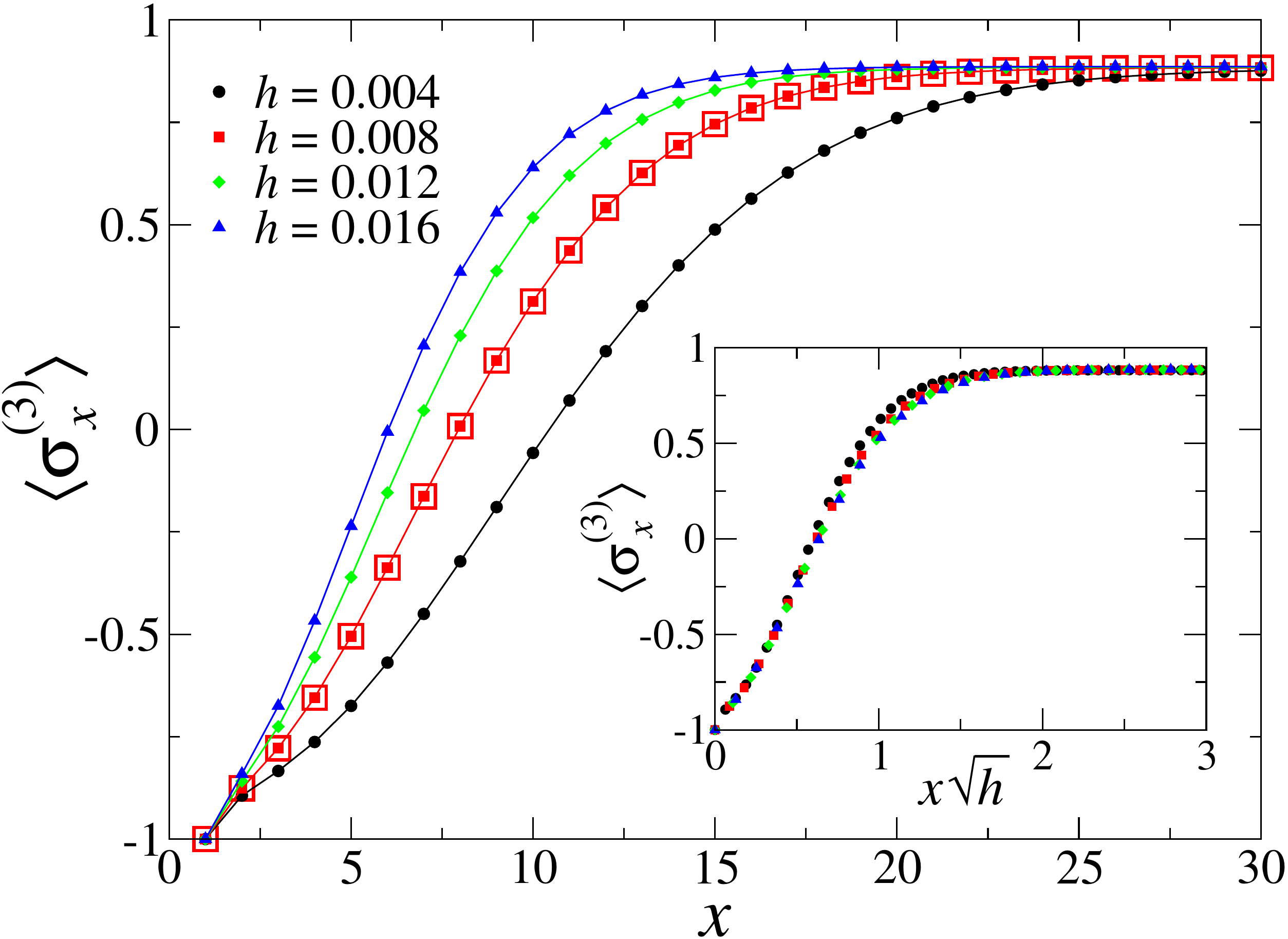}
  \caption{Magnetization profile close to the boundaries, for $g=0.8$
    and several values of $h$, as indicated in the legend.  Data are
    all for $L=180$ (filled symbols), except empty red squares, which
    stand for $L=100$ and $h=0.008$ and are superposed to filled red
    squares (on the scale of the figure).  The inset shows data
    collapse, after a rescaling of the $x$ position by a factor
    $\sqrt{h}$.}
  \label{fig:bouproMx}
\end{figure}

\begin{figure}[!t]
  \includegraphics[width=0.95\columnwidth]{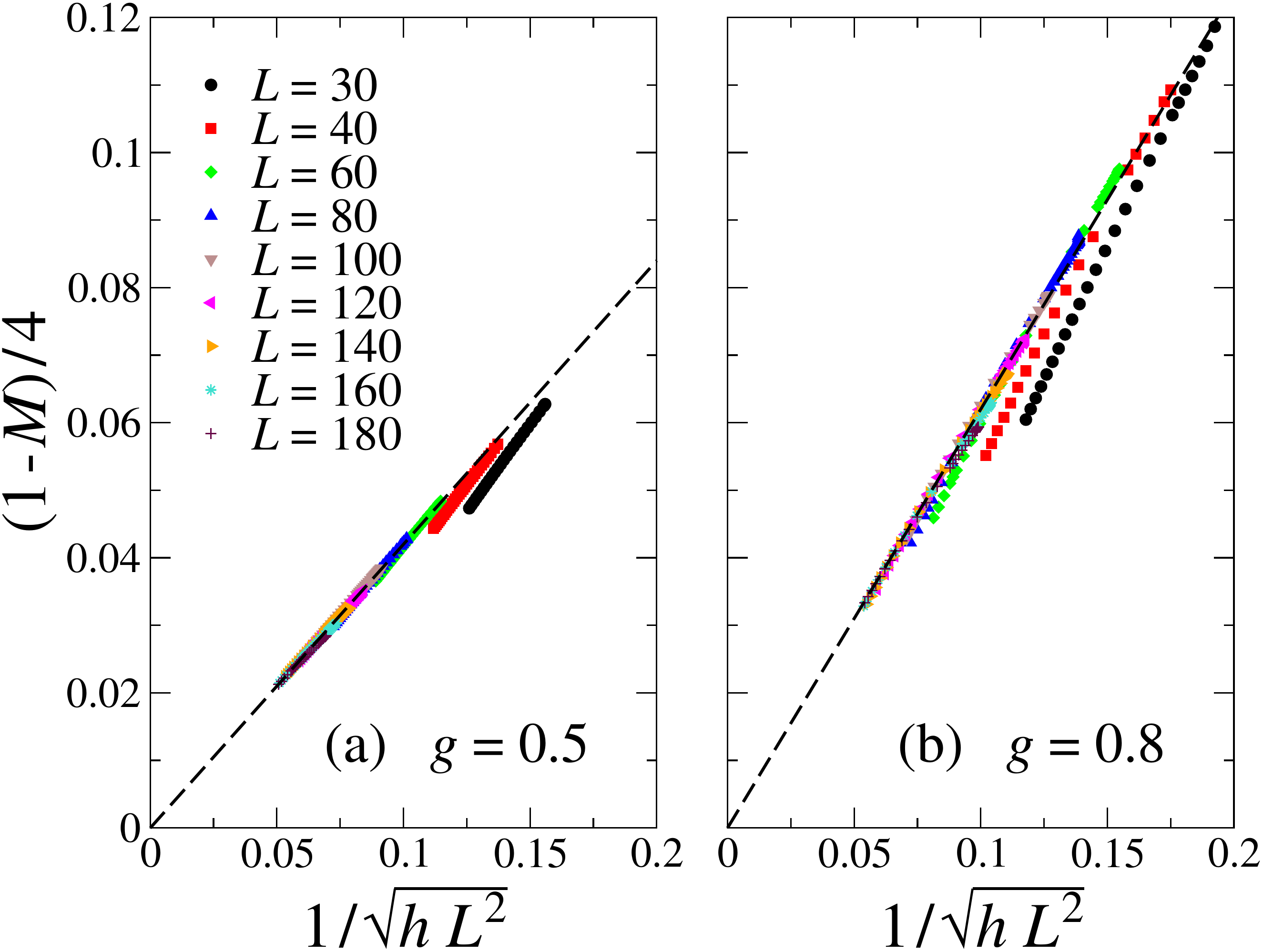}
  \caption{Relative length $v_- = (1-M)/4$ of the negatively
    magnetized phase, as a function of the rescaled variable
    $1/(h^{1/2}L)$.  The various data sets are for different system
    sizes $L$, while the two panels are for $g=0.5$ (a) and for
    $g=0.8$ (b).  Dashed straight lines are constrained fits of the
    numerical data at large $L$ to Eq.~\eqref{Msca}.}
  \label{fig:largehaveM}
\end{figure}

In Fig.~\ref{fig:bouproMx} we report the magnetization profile close
to one of the boundaries. The local magnetization differs from the
value at the center, in a region of size $\ell_-$ close to $x=0$.  The
region where $M$ varies significantly shrinks as $h$ increases, as
expected. A detailed analysis of the data shows that $\ell_-$ has a
non trivial power-law dependence on $h$, for $h$ large enough.
Indeed, numerical results and a phenomenological theory for the
magnetization profile in this regime --- see Sec.~\ref{phenth} --- lead us
to conjecture that
\begin{equation}
\ell_- \sim h^{-1/2} .
\label{elldeph}
\end{equation}
The emergence of this behavior is clearly supported by the plot
reported in the inset of Fig.~\ref{fig:bouproMx}. If we plot the
magnetization data versus $x \sqrt{h}$, we observe the collapse of the
data.  This is consistent with the data for a wide range of values of
$L$ and $h$, and for two different values of $g$.  The scaling
\eqref{elldeph} implies that the relative size of the negatively
magnetized region at one of the boundaries behaves as
\begin{equation}
v_- \equiv {\ell_-\over L} \sim  {1 \over h^{1/2} L}.
\label{v-beh}
\end{equation}
Equation (\ref{v-beh}) implies a scaling behavior for the average
magnetization $M$, since this quantity is sensitive to the behavior at
the boundaries. If we make a simple approximation in which the
magnetization is $-1$ in two boundary regions of linear size $\ell_-$
and $+1$ in the central region of linear size $L-2\ell_-$, we predict
$M = 1 - 4 \ell_-/L = 1 - 4 v_-$. We can take this equation as the
definition of the relative length of the region in which the
magnetization is negative.  Equation~(\ref{v-beh}) then predicts
\begin{equation}
v_- = {1-M\over 4} \approx {a(g)\over h^{1/2} \, L},
\label{Msca}
\end{equation}
for $h>h_{tr}(L)$.  This scaling behavior is clearly supported by the
data displayed in Fig.~\ref{fig:largehaveM}, for two different values
of $g$. We estimate $a(g=0.5)\approx 0.42$ and $a(g=0.8)\approx 0.62$
(see dashed straight lines in the figure).  Therefore,
Eq.~(\ref{v-beh}) signals the presence of two negatively magnetized
regions, whose width widens as $h$ decreases at fixed $L$. However,
since the scaling applies only up to $h_{tr}(L)$, at fixed $L$, the
width $v_-$ satisfies
\begin{equation}
   v_- \le v_{\rm max} = {a(g) \over h_{tr}(L)^{1/2} \, L} \sim L^{-1/2},
\end{equation}
i.e., the maximum relative size decreases with $L$. Note that $v_-$
decreases with increasing $L$ only outside the transition region close
to $h_{tr}(L)$.  In the scaling region around $h_{tr}(L)$, $v_-$
remains finite as $L$ increases.

\begin{figure}[!t]
  \includegraphics[width=0.95\columnwidth]{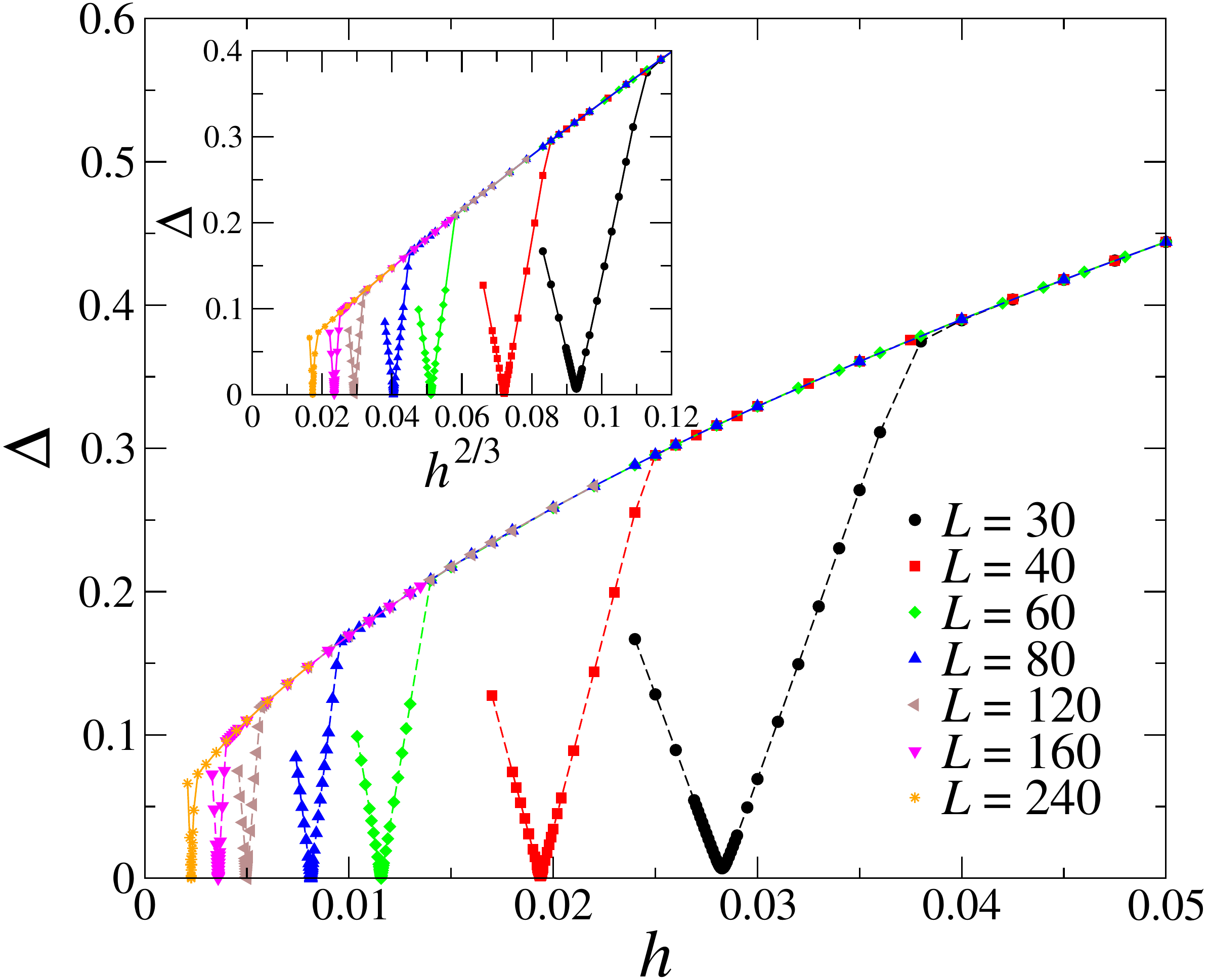}
  \caption{Energy gap $\Delta(L,h)$ versus $h$ in the large-$h$
    region, for $g=0.8$ and different system sizes, as indicated in
    the legend.  The inset displays the same data versus
    $h^{2/3}$. The dependence $\Delta \sim h^{2/3}$, see
    Eq.~\eqref{largehdeka}, emerges quite clearly.}
  \label{fig:largehDelta}
\end{figure}

Let us finally analyze the energy gap $\Delta(L,h)$ between the two
lowest-energy states.  Results are reported in
Fig.~\ref{fig:largehDelta}.  For each value of $L$, the gap shows
three distinct behaviors.  For small magnetic fields $\Delta$
decreases. At $h_{tr}(L)$ it is essentially zero on the scale of the
figure, then it increases sharply up to an $L$-dependent value
$h_\times (L)$~\cite{footnotehtimes}.  Finally, for $h > h_\times (L)$
it follows an $L$-independent curve. In the latter regime, the gap
behaves as
\begin{equation}
\Delta(L,h) \sim h^{2/3},
\label{largehdeka}
\end{equation}
as it appears neatly from the rescaling provided in the inset.  Note
that Eq.~(\ref{largehdeka}) is the expected behavior for kink-antikink
states in an external longitudinal magnetic field --- see, e.g.,
Refs.~\cite{MW-78,Coldea-etal-10,Rutkevich-10}.  This confirms our
conjecture that the general features of the low-energy properties for
$h>h_{tr}$ are related to states with positively and negatively
magnetized regions separated by kink-like structures.

The scaling (\ref{largehdeka}) explains the nonanalytic behavior of
$h_{\rm tr}(L)$, outlined in Eq.~(\ref{fithl}).  Indeed, assume that
the negatively magnetized state has an energy that scales as $E_{\rm
  magn} \approx E_0(L) - a_m h L - b_m h$, while all kink-antikink
states have an energy that, consistently with the result
(\ref{largehdeka}), scales as $E_{\rm kink} = E_1(L) + a_k h L + c_k
h^{2/3} + b_k h$.  Equating the two energies, $E_{\rm magn} = E_{\rm
  kink}$, and taking into account that $E_1(L) - E_0(L)$ is finite for
large $L$, we obtain the behavior (\ref{fithl}), and in particular the
$O(L^{-\zeta})$ correction with $\zeta=2/3$.

\subsection{Phenomenological theory for the large-$h$ region}
\label{phenth}

We develop here a phenomenogical theory for the behavior of the
system.  Given the numerical results, two states are relevant for the
system. One should consider the negatively magnetized state
$\mu(x) = -1$ of energy $E = h L$ (hereafter we adopt the shorthand
notation $\mu(x) \equiv \langle \sigma^{(3)}_x \rangle$),
which has average renormalized magnetization $M = -1$.
The second relevant state is a double-kink state,
characterized by two boundary regions of size $\xi$ in which the
magnetization is less than 1, and by one central region of size
$L-2\xi$ in which $\mu(x) = 1$. The average magnetization of the
double-kink state is
\begin{equation}
M = {1\over L} \left( L - {a_1 \xi} \right) ,
\end{equation}
where $a_1$ is an appropriate constant. Correspondingly, its energy is
\begin{equation}
E = E_0 - h L M = E_0 - h L + a_1 \xi h,
\end{equation}
where $E_0$ is the energy of the state in the absence of magnetic
field.  To compute $E_0$ we use a phenomenological approach.  We
assume that, for $h = 0$, the Hamiltonian can be written in terms of
$\mu(x)$ as
\begin{equation}
H = \int_0^L dx\,
 \left[ a \left( {d \mu(x) \over dx}\right)^2 + b \Big( \mu(x)^2 - 1 \Big)^2 \right].
\end{equation}
Boundary conditions require $\mu(0) = -1$ and $\mu(L) = -1$.
The double-kink state can be  parametrized as 
\begin{equation}
  \mu(x) = A \tanh [B x] + A \tanh [B (L-x)] + C , 
\label{Mx-dk}
\end{equation}
where $C$ should be fixed to guarantee the boundary conditions.
Equation~(\ref{Mx-dk}) holds only if the localized kink and antikink do not
interact, which in turn requires that $BL\gg 1$. If we further assume
that $\mu(L/2) = 1$, the profile can be written as
\begin{equation}
\mu(x) = 2 \tanh [B x] + 2 \tanh [B (L-x)] - 3 .
\label{Mx-dk2}
\end{equation}
It is clear that $B$ should be identified with the parameter $\xi$ 
defined before.
A simple computation gives  
\begin{eqnarray}
H = {1\over B} a_0 + a_2 B \qquad {\rm for}\;BL\gg 1, 
\end{eqnarray}
with 
\begin{equation}
a_0 = {64\over 3} (3 \ln 2 - 2)b, \qquad
a_2 = {16\over 3} a.
\end{equation}
Therefore, the energy of the double-kink state in a magnetic field is 
\begin{equation}
E = a_0 \xi + {a_2\over \xi} - h L + h a_1 \xi.
\end{equation}
The ground-state energy is obtained by minimizing $E$ with respect to $\xi$.
We obtain
\begin{equation}
\xi = \left({a_0 + a_1 h\over a_2}\right)^{-1/2},
\end{equation}
and, correspondingly, 
\begin{eqnarray}
E & = & 2 \sqrt{a_2(a_0 + a_1 h)} - h L,  \\
M & = & 1 - {a_1\over L} \left( {a_2 \over a_0 + a_1 h} \right)^{1/2}. 
\end{eqnarray}
The double-kink state competes with the magnetized one with
$\mu(x) = -1$.
The transition between the corresponding large-$h$ and small-$h$
regimes, where the ground states are the double-kink and magnetized
states respectively, occurs for
\begin{equation}
h = h_{tr} = {(a_0 a_2)^{1/2} \over L}.
\label{htrmod}
\end{equation}
Close to $h_{tr}$, $\xi$ is a finite number.  The behavior changes as
$h$ increases. If $a_1 h \gg a_0$ we find $\xi \sim h^{-1/2}$ which
shows that the region in which $\mu(x) < 1$ shrinks as
$h^{-1/2}$. Moreover,
\begin{eqnarray}
H & \approx & - h L\left[1 - 2 \left({a_1 a_2 \over h L^2}\right)^{1/2}\right], \\
M & \approx & 1 - \left({a_1 a_2 \over h L^2}\right)^{1/2}.
\end{eqnarray}
Corrections to scaling are functions of $h L^2$, in agreement with numerical
results --- see Fig.~\ref{fig:largehaveM}.

\subsection{Localized magnetic field}
\label{locmegfi}

It is likewise interesting to discuss the case of an Ising chain with
a localized magnetic field.  To this purpose, let us consider a chain
with an odd number of sites, $L= 2\ell + 1$, whose Hamiltonian is
obtained by replacing the homogeneous term $- h \,\sum_{x=1}^L
\sigma^{(3)}_x$ with a local term $- h_l \sigma^{(3)}_{x_c}$ in
Eq.~(\ref{hedef2}), where $x_c$ is the central site. As in the
homogeneous case, we can easily identify two distinct regions, in
which ground-state properties are different. For small $h_l$, the
system is magnetized, as before.  For large $h_l$, it is enough to
observe that, in the ground state, the central site is essentially
fixed, as it should be aligned with the magnetic field. Therefore, the
ground state is equivalent to that of two disjoint chains with fixed
and opposite boundary conditions.  Using the results of
Refs.~\cite{CNPV-14,CPV-15b}, we can conclude that the ground state is
a kink state, with zero average magnetization. In this case the gap
is~\cite{CPV-15b}
\begin{equation}
  \Delta(\ell) \approx {3 g\over 1-g} \, {\pi^2\over \ell^2}\,. 
  \label{delell}
\end{equation}
Thus, for $h_l$ small and $h_l$ large, the nature of the ground state
differs.  Therefore, we expect two different regions: a magnetized
region for $h < h_{l,tr}$ and a kink phase for $h > h_{l,tr}$.  For
$h_l=h_{l,tr}(L)$ a sharp transition occurs between the magnet and
kink phases~\cite{CPV-15b}, where the magnetization profile is
expected to qualitatively change.  Its location $h_{l,tr}(L)$ is
expected to be associated with the minimum $\Delta_m(L)$ of the gap
$\Delta(L,h_l)$, i.e., $\Delta_m(L)\equiv\Delta[L,h_{l,tr}(L)]$.

\begin{figure}[!t]
\includegraphics[width=0.95\columnwidth]{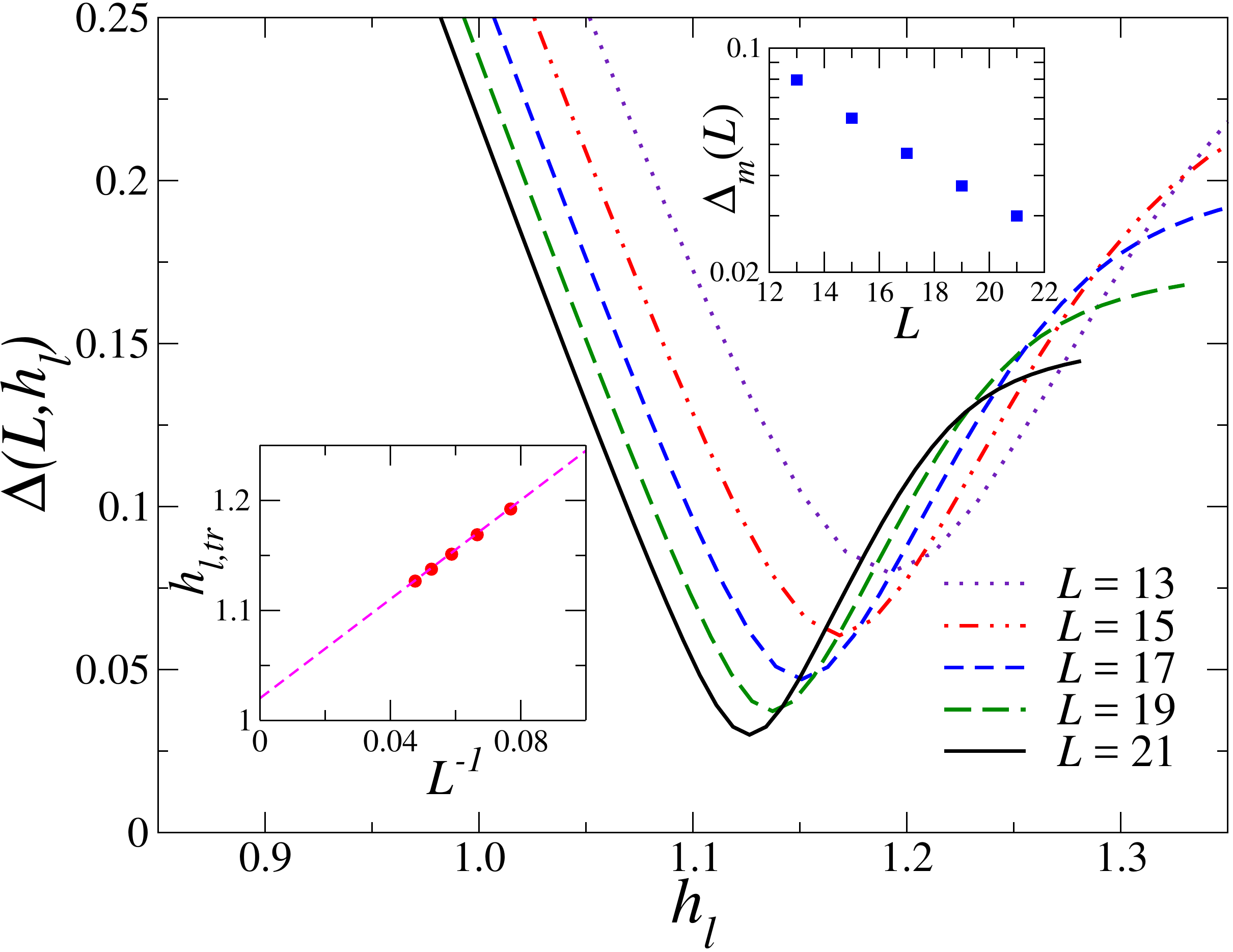}
  \caption{The energy difference (gap) of the lowest states as a
    function of the localized magnetic field $h_l$ at the center of
    the chain, for $g=0.5$.  The insets show its minimum $\Delta_m(L)$
    (top inset) and the corresponding $h_{l,tr}(L)$ (bottom inset),
    which appears to approach a constant value with $O(L^{-1})$
    corrections.  Indeed, by fitting the available data to
    $h_{l,tr}(L) = h_{l,tr}^* + b/L$, we obtain $h_{l,tr}^* = 1.02$,
    see the dashed line in the corresponding inset.  }
  \label{fig:deltaloc}
\end{figure}

\begin{figure}[!t]
\includegraphics[width=0.95\columnwidth]{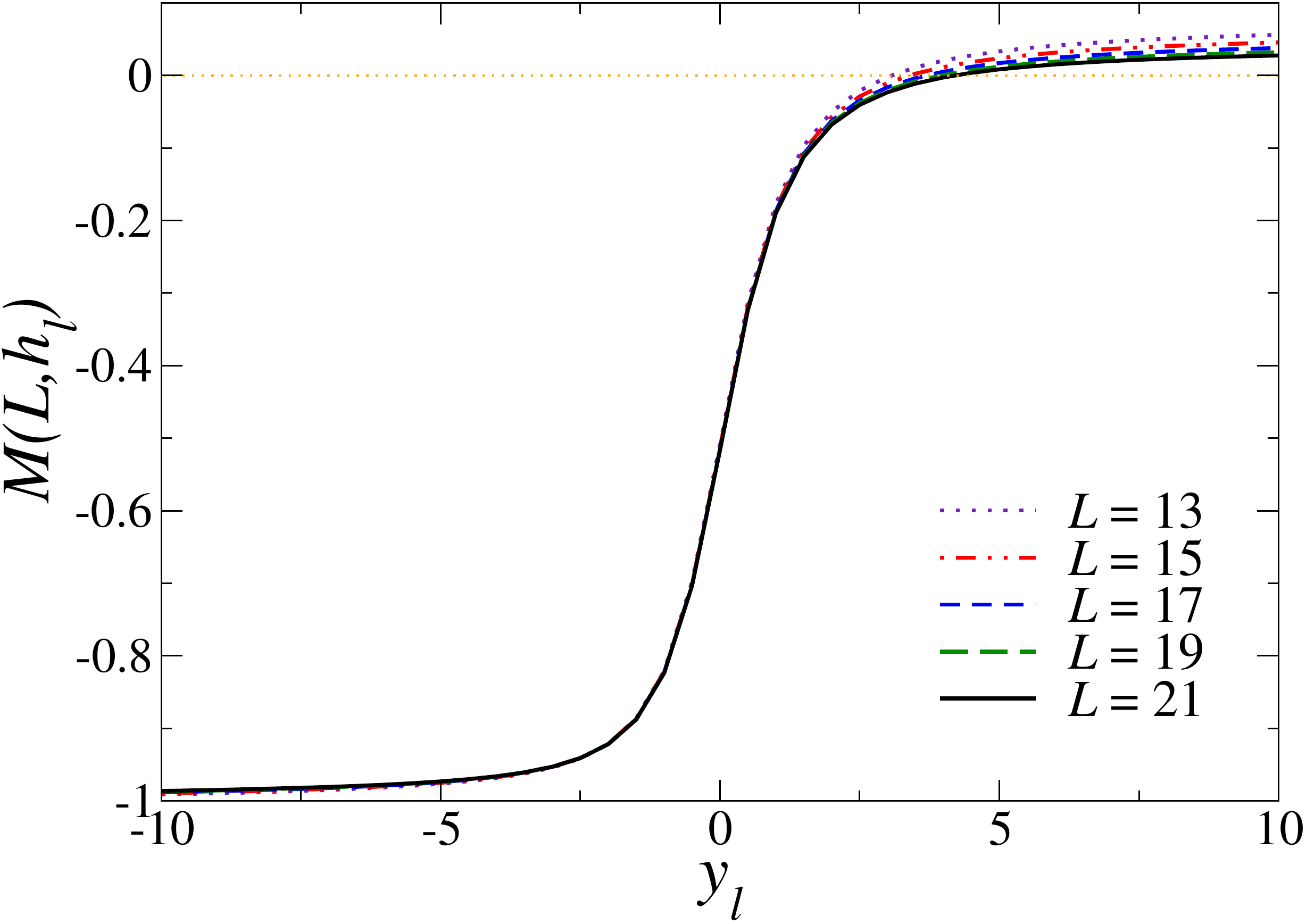}
  \caption{Renormalized average magnetization $M(L,h_l)$ for $g=0.5$
    versus $y_l$ [see Eq.~(\ref{def-yloc})] for the model with a local
    magnetic field at the center of the chain. The curves appear to
    approach a scaling function ${\cal M}_l(y_l)$ with increasing $L$,
    i.e. $M(L\to\infty,h_l)\approx {\cal M}_l(y_l)$. Note that,
    although the magnetization appears positive for $y_l>0$, its
    large-$L$ limit is consistent with negative extrapolations, and
    ${\cal M}_l(y_l\to\infty)=0$.}
  \label{fig:Mhloc}
\end{figure}

This picture is confirmed by the numerical data, as shown in
Fig.~\ref{fig:deltaloc} for $g=0.5$.  As expected, $h_{l,tr}(L)$
converges to a finite value for $L\to \infty$. For $h_l$ close to such
value, the relevant FSS variable is expected to be
\begin{equation}
y_l = {2 m_0 [h_l - h_{l,tr}(L)] \over \Delta_{m}(L)}.
\label{def-yloc}
\end{equation}
We note that, like for the case of the global magnetic field considered
previously, the definition of the scaling variable $y_l$ requires the
actual values of $h_{l,tr}(L)$ and $\Delta_m(L)$ for the size $L$, and not
their asymptotic behaviors.  Then, in the large-$L$ limit, we expect
that
\begin{equation}
 M(L,h)\approx {\cal M}_l(y_l).
\label{locma}
\end{equation} 
The scaling function ${\cal M}_l(y_l)$ is expected to be negative, and
asymptotically ${\cal M}_l(y_l\to\infty) = 0$.  The numerical data in
Fig.~\ref{fig:Mhloc} clearly support this scaling behavior.
Qualitatively, we therefore obtain the same behavior as in the
homogeneous case.  Quantitatively, however, there are important
differences.  For instance, in the large-$h_l$ region (i.e., for $h_l
\gg h_{l,tr}$), the average magnetization is smaller than 1 and
correspondingly the size $\ell_-$ of the negatively magnetized region
at the boundaries is of order $L$, and not of order $L^{1/2}$.

\subsection{Summary}
\label{summaryssec}  

Our numerical results show that the quantum Ising chain with EFBC
favoring one of the magnetized phases develops notable scaling
features along the FOQT line.  The FSS arising from the interplay
between the size $L$ and the bulk longitudinal field $h$ turns out to
be intriguingly more complex than that observed with neutral boundary
conditions, see in particular Sec.~\ref{equiopbc} for PBC and OBC.  In
the case of EFBC with both ends favoring the same phase, the
observables around $h=0$ depend smoothly on $h$, up to a pseudo
transition value $h_{tr}(L)$, behaving as $h_{tr}(L)\approx \eta/L$,
where they develop a singularity. This corresponds to a sharp
transition to the oppositely magnetized phase, which appears to be
analogous to a discontinuous transition. Around the finite-size
transition point $h_{tr}(L)$, the system develops a FSS 
controlled by an exponentially vanishing gap $\Delta(L)\sim e^{-bL}$.
The scaling arises from the competition of the two lowest-energy
states, indeed, scaling functions correspond to those of a two-level
system. The relevant low-energy states are superpositions of a
negatively magnetized state, which is the ground state for
$h<h_{tr}(L)$, and of a state with a positively magnetized region
around the center and two negatively magnetized regions at the
boundaries, separated by kink-like structures (see the magnetization
profile shown in Fig.~\ref{fig:Mprofile}).  The latter state becomes
the ground state for $h>h_{tr}$. Outside the transition region, it is
characterized by two negatively magnetized regions close to the
boundaries, of typical size $\ell_-\sim h^{-1/2}$.

The classical counterpart of this complex scenario for quantum
many-body systems at FOQTs has been investigated at thermal
first-order transitions in statistical systems with disordered
boundary conditions~\cite{PPV-18}, where the interplay between the
temperature and the finite size gives rise to a complex scenario as
well, characterized by different scaling regions.

We also considered the case of a localized external longitudinal
field.  Again we can identify two different regimes separated by a
transition at a finite value of the local magnetic field. However, at
variance with what happens in the case of PBC and OBC (see the
discussion in Sec.~\ref{equiopbc}), the nature of the transition and of
the high local-field phase differs. In particular, its FSS at the
transition does not arise from an avoided two-level crossing 
 in finite systems, and the average magnetization cannot
exceed $M=0$.

\section{Conclusions}
\label{conclu}

We have investigated FSS at FOQTs when boundary conditions favor one
of the two phases.  We have shown that substantial differences emerge
with respect to neutral boundary conditions, such as PBC.

For this purpose, we presented a numerical study of one of the
simplest paradigmatic quantum many-body systems exhibiting a
nontrivial zero-temperature behavior: the one-dimensional quantum
Ising chain in the presence of a transverse field, whose
zero-temperature phase diagram features a line of FOQTs driven by a
longitudinal external field.  We provided evidence that the interplay
between the size $L$ and the bulk longitudinal field $h$ is more
complex than that observed with neutral boundary conditions.  In the
case of EFBC favoring the same phase, for small values of $h$,
observables depend smoothly on $h$, up to $h_{tr}(L)\approx \eta/L$,
where $\eta$ is a $g$-dependent constant, where a sharp transition to
the oppositely magnetized phase occurs.  In proximity of $h_{tr}(L)$,
a universal FSS behavior emerges from the competition of the two
lowest-energy states, separated by a gap which vanishes exponentially
with $L$.  For even larger longitudinal fields, $h>h_{tr}(L)$, a scaling
behavior controlled by another variable $\sim 1/( h^{1/2} \, L)$ appears.

We believe that analogous behaviors can be observed in other FOQTs,
when boundary conditions favor one of the two phases. In particular,
they should also occur in systems defined in more than one
dimension. It would be also interesting to verify whether these
complex behaviors may occur in FOQTs driven by even perturbations as well,
where OBC favor the disordered phase and EFBC the ordered one. Some
results for the quantum Potts chain appeared in Ref.~\cite{CNPV-15},
where, however, only the behavior around $h=0$, analogous to that
discussed in Sec.~\ref{smallh}, was considered.

Finally we mention that it would be tempting to generalize the
discussion to off-equilibrium dynamics across the FOQT.  For example,
one could address a situation where the longitudinal field is subject
to a time-dependent driving, and devise suitable scaling laws which
may depend on the properties of the equilibrium
transition~\cite{CEGS-12}, in analogy to what has been done so far for
the same system with neutral boundary conditions~\cite{PRV-18}.
However, in view of the presence of several different equilibrium
scaling behaviors, the off-equilibrium dynamics in the presence of
EFBC may exhibit an intriguing, and possibly more complex, scenario.

Quite remarkably, the FSS behavior outlined in this paper can be
observed for relatively small sizes: in some cases a limited number of
spins already displays the asymptotic behavior.  Therefore, even
systems of modest size ($L \lesssim 100$) may show definite signatures
of the scaling laws derived in this work.  In this respect,
present-day quantum-simulation platforms have already demonstrated
their capability to reproduce and control the dynamics of quantum
Ising-like chains with a small number of spins.  Ultracold atoms in
optical lattices~\cite{Bloch-08, Simon-etal-11}, trapped
ions~\cite{Edwards-etal-10,Islam-etal-11,LMD-11,Kim-etal-11,Debnath-etal-16},
and Rydberg atoms~\cite{Labuhn-etal-16} seem to be the most promising
candidates where the emerging universality properties of the quantum
many-body physics discussed here can be tested with a minimal number
of controllable objects.


\begin{thebibliography}{99}
\bibitem{footnote} Since we are interested in the ground-state
  physics, the full diagonalization of the Hamiltonian matrix is not
  required. For this reason, we implemented a much faster Lanczos
  diagonalization technique for systems with up to $L=22$ sites,
  corresponding to a Hilbert space of dimension $2^{22} \sim 4 \times
  10^6$.  For larger sizes, a standard finite-size DMRG algorithm
  enabled us to reach systems with few hundreds of sites using a
  limited amount of resources.  We kept the truncation error below
  $10^{-8}$, using a number of states up to $m \approx 400$.
\bibitem{footnotehtimes}
  In the limit $g \to 0$, we can easily verify that $h_\times(L)$ scales 
  as $1/L$. Our results for finite values of $g$ are also consistent with this
  behavior.
\end{thebibliography}
\end{document}